\documentclass[a4paper,english,aps,twocolumn,floatfix,amsfonts,amssymb,superscriptaddress]{revtex4}
\usepackage{epstopdf}
\usepackage{graphicx}
\usepackage{dcolumn}
\usepackage{bm,bbm,bbold}
\usepackage{color,xcolor, soul}
\sethlcolor{green}
\usepackage{amsfonts}
\usepackage{amsmath}
\usepackage{mdframed}
\usepackage[normalem]{ulem}
\usepackage{mathrsfs}   
\usepackage[percent]{overpic}
\usepackage[colorlinks=true,citecolor=blue]{hyperref}
\usepackage{hyperref}
\hypersetup{colorlinks=true,citecolor=blue,linkcolor=blue,urlcolor=blue}
\begin{document}
\title{Piezoelectricity and valley Chern number in inhomogeneous hexagonal 2D crystals}
\author{Habib Rostami}
\email{habib.rostami@su.se}
\affiliation{Istituto Italiano di Tecnologia, Graphene Labs, Via Morego 30, I-16163 Genova,~Italy}
\author{Francisco Guinea}
\affiliation{Fundaci\'on IMDEA Nanociencia, C/Faraday 9, Campus Cantoblanco, 28049 Madrid, Spain}
\affiliation{School of Physics and Astronomy, University of Manchester, Oxford Road, Manchester M13 9PL, UK}
\author{Marco Polini}
\affiliation{Istituto Italiano di Tecnologia, Graphene Labs, Via Morego 30, I-16163 Genova,~Italy}
\affiliation{School of Physics and Astronomy, University of Manchester, Oxford Road, Manchester M13 9PL, UK}
\author{Rafael Rold\'an}
\email{rroldan@icmm.csic.es}
\affiliation{Instituto de Ciencia de Materiales de Madrid, CSIC, 28049 Cantoblanco, Madrid, Spain}
\begin{abstract}Conversion of mechanical forces to electric signal is possible in non-centrosymmetric materials due to linear piezoelectricity. 
The extraordinary mechanical properties of two-dimensional materials and their high crystallinity make them exceptional platforms to study and exploit the piezoelectric effect.
Here, the piezoelectric response of non-centrosymmetric hexagonal two-dimensional crystals is studied using the modern theory of polarization and ${\bm k} \cdot {\bm p}$ model Hamiltonians.  An analytical expression for the piezoelectric constant is obtained in terms of topological quantities such as the {\it valley Chern number}. The theory is applied to semiconducting transition metal dichalcogenides and hexagonal Boron Nitride. We find good agreement with available experimental measurements for MoS$_2$. We further generalise the theory to study the polarization of samples subjected to inhomogeneous strain (e.g.~nanobubbles). We obtain a simple expression in terms of the strain tensor, and show that charge densities $\gtrsim 10^{11} {\rm cm}^{-2}$ can be induced by realistic inhomogeneous strains, $\epsilon \approx 0.01 - 0.03$.
\end{abstract}
\date{\today}
\maketitle
{\it Introduction.---} 
Piezoelectricity is a property of crystals with broken inversion symmetry, which allows conversion of mechanical to electric energy~\cite{Brown_1962,Ramadan_2014}. When subjected to an external strain field ${\bm \varepsilon}$, piezoelectric crystals acquire a polarization ${\bm P}$ that is described by the third-rank piezoelectric tensor $\gamma_{ijk} \equiv \partial P_i/\partial \varepsilon_{jk}|_{{\bm \varepsilon}\to {\bm 0}}$. The so-called {\it modern theory of polarization} exploits the properties of the Berry connection (BC) of the electronic wave-functions to quantify the change of polarization in an extended system~\cite{KS_DV_1993,Resta_1994,Resta_Vanderbilt,Resta_2010}. For crystalline insulators, the BC is obtained in terms of the Bloch orbitals, and the polarization can thus be calculated as an integral of the BC on whole Brillouin zone. This quantum mechanical description of the polarization has been used to calculate the piezoelectric constant of a number of crystals from {\it ab initio} \cite{Duerloo_2012,Michel_2017} as well as analytical approaches \cite{Mele_PRL_2002,Droth_2016}.
\par
Inversion symmetry is broken in a large number of two-dimensional (2D) materials \cite{Roldan_CSR_2017}. This, together with their exceptional breaking strength and flexibility  
\cite{Lee385,Roldan_JPCM_2015}, make them perfect platforms for strain engineering and, in particular, for piezoelectric applications~\cite{Wang2010}.  Indeed, the isolation of monolayer and few-layer crystals of transition metal dichalcogenides (TMDs) or hexagonal Boron Nitride (h-BN)~\cite{Novoselov_PNAS_2005} provides materials with symmetry properties that are different from their bulk counterparts. Bulk TMDs with the common formula $MX_2$ ($M=$~Mo,W and $X=$~S,Se) consist of stacked layers of $MX_2$ monolayers bonded by van der Waals forces, and have a centre of symmetry located between the layers. Therefore bulk TMDs are not piezoelectric. However, isolation of a monolayer of $MX_2$ from the bulk crystal removes the center of symmetry, leading to piezoelectricity, as reported experimentally~\cite{Wu_2014,Zhu_2015}. Similarly, h-BN consists of a honeycomb lattice with different elements in the two sublattices of the unit cell, making this material piezoelectric as well.  
Both monolayer h-BN and TMDs are hexagonal crystals that belong to the $D_{3h}$ point symmetry group which contains two main symmetry elements: mirror reflection $\sigma_{\rm v}:x\to-x$, and three-fold $C_{3}$ rotational symmetries, with the $\hat {\bm x}$ axis along the zigzag direction. They present a direct band gap at the two inequivalent K and K$'$ points of the Brillouin zone (BZ), and their low-energy electronic excitations are well described by massive Dirac-like Hamiltonians~\cite{Fuchs_EPJB_2010,Xiao2012,Rostami_prb_2013}. 
\par
Realistic samples are often subject to non-uniform strain. This is particularly common in 2D crystals which are strained by controlled corrugation \cite{Castellanos-Gomez_NL_2013,Quereda_NL_2016}, deposition on substrates with nanodomes \cite{Li_NC_2015} or nanopillars \cite{Branny_2017,Palacios-Berraquero_2017}, or because of the formation of bubbles due to trapped substances between the 2D crystal and the substrate \cite{Khestanova_NC_2016}. 
If the crystal is non-centrosymmetric, non-uniform strain can be a source of carrier doping, with charge density given by $\rho({\bm r})=-{\bm \nabla}\cdot{\bm P}({\bm r})$~\cite{Griffiths}, due to local variations of the polarization. This issue will be one of the main focuses of this work. 
\par
Using a generic ${\bm k} \cdot {\bm p}$ model Hamiltonian, we derive analytical expressions for the piezoelectric coefficients of hexagonal 2D crystals. Explicit calculations for TMDs (MoS$_2$, MoSe$_2$, WS$_2$ and WSe$_2$) and h-BN are reported. Good agreement is found in comparison to existing {\it ab initio} calculations and experimental measurements of piezoelectric constant. We further study the strain-induced polarization in undoped samples subject to non-uniform strain, like Gaussian and triangular bumps, bubbles, etc., finding that inhomogeneous deformations can induce large charge densities. 
TMDs are being extensively studied as platforms where the gap can be locally manipulated by strain \cite{Cetal13,Metal15,Li_NC_2015,Letal16}, and where strain give rises to optical single-photon sources \cite{KKG15,Branny_2017,Palacios-Berraquero_2017} (quantum emitters). Our theory can be used to determine the charge densities induced in these systems, as function of strain and device size.
\par
{\it General formulation.---}
Let us consider a 2D crystal subject to a uniform static strain field. In the linear response regime, the induced-polarization due to the piezoelectric effect is given by
$P_{i} = \sum_{jk}\gamma_{ijk}\varepsilon_{jk} $ where $\gamma_{ijk}$ and $\varepsilon_{jk}$ are the piezoelectric and the strain tensors, respectively. The quantity $\gamma_{ijk}$ must respect the symmetries of the lattice, implying $\sum_{i' j' k'} R^\dagger_{i i'} \gamma_{i' j' k' } R_{j' j} R_{k' k} = \gamma_{ijk}$ where ${\bm R}$ accounts for a point group symmetry element. For instance, for a 2D system with $D_{3h}$ symmetry lying in the $xy$-plane, after considering $\sigma_{\rm v}$ and $C_{3}$ symmetries, we find that $\gamma_{xxx}=\gamma_{xyy}=\gamma_{yxy}=\gamma_{yyx} =0$, and $\gamma_{yyy} = -\gamma_{yxx}=-\gamma_{xyx} =-\gamma_{xxy}$~. The above symmetry properties allow us to write the piezoelectric-induced polarization as (see Appendix \ref{app:sym}) 
\begin{equation}\label{eq:P_sym}
{\bm P}= \gamma_{yyy} { \bm {\mathcal A}} \times \hat {\bm z}~,
\end{equation}
where ${ \bm {\mathcal A}} =  (\varepsilon_{xx}-\varepsilon_{yy}) \hat{\bm x} - 2 \varepsilon_{xy}  \hat{\bm y}$. It is worth noting that this expression is formally equivalent to the gauge field that describes the effect of strains on the electronic structure of graphene \cite{Guinea_2010}.
Notice also that, according to Eq.~(\ref{eq:P_sym}), the charge polarization is always perpendicular to $ { \bm {\mathcal A}}$, a result that has been reported in Ref.~\cite{Droth_2016}. 
Finally, Eq.~(\ref{eq:P_sym}) also implies that the trace of the strain tensor, $V=\varepsilon_{xx}+\varepsilon_{yy}$, does not contribute to the polarization. 
From now on, we use $\gamma$ to indicate $\gamma_{yyy}$.
According to the  modern theory of polarization \cite{KS_DV_1993,Resta_1994,Resta_Vanderbilt,Resta_2010}, the electronic polarization of an insulator, ${\bm P}$, can be calculated from the geometrical properties of the Bloch wave-functions, 
\begin{equation}\label{eq:p_el}
{\bm P}({ \bm {\mathcal A}})  = -e \sum_{\tau s} \int_{\rm BZ} \frac{d {\bm k} }{(2\pi)^2}  { \bm a}^{({\rm v})}_{\tau s}({\bm k}, \tau { \bm {\mathcal A}})~,
\end{equation}
where $s=\pm$ and $\tau=\pm$ account for the spin and valley degrees of freedom, respectively.  The valence-band BC reads ${ \bm a}^{({\rm v})}_{\tau s}({\bm k},\tau { \bm {\mathcal A}}) = i \langle u^{({\rm v})}_{\tau s} ({\bm k},\tau { \bm {\mathcal A}}) | {\bm \nabla}_{\bm k} | u^{({\rm v})}_{\tau s}({\bm k},\tau { \bm {\mathcal A}})   \rangle$ where $  | u^{({\rm v})}_{\tau s}({\bm k},\tau { \bm {\mathcal A}})\rangle$ is the eigenvector of the system under strain. The piezoelectric constant is obtained as (see Appendix \ref{app:piezo_const})
\begin{equation}
\gamma = \frac{e}{2} \sum_{\tau, s} \int_{\rm BZ} \frac{d {\bm k} }{(2\pi)^2} 
\lim_{\small \bm {\mathcal A} \to {\bm 0}}\left [  \frac {\partial a^{({\rm v})}_{\tau s,y}({\bm k}, \tau { \bm {\mathcal A}})}{\partial {\cal A}_x} - (x \leftrightarrow y) \right].
\end{equation}
In the linear response regime with respect to the strain gauge field, we can formally  write the electronic Hamiltonian as ${\cal H}_{\tau s}({\bm k},\tau{ \bm {\mathcal A}}) \approx {\cal H}_{\tau s}({\bm k}) + \tau \sum_{\alpha}  {\cal  A}_\alpha \partial {\cal H}_{\tau s}({\bm k}, { \bm {\mathcal A}})/\partial {\cal A}_\alpha |_{\small {\cal A} \to 0}+ {\cal O}({\cal A}^2) $, where $ {\cal H}_{\tau s}({\bm k})$ is the unstrained Hamiltonian. Moreover, the valence band eigenvector can be evaluated by using first-order perturbation theory. After some straightforward calculations, we obtain the following expression for the piezoelectric coefficient
\begin{equation}\label{Eq:gamma-Chern}
\gamma = \frac{e}{4\pi a_0} \sum_{\tau s} \tau \widetilde{\cal C}_{\tau s}~,
\end{equation}
where $a_0$ is an effective lattice constant, $\widetilde{\cal C}_{\tau s} = \int_{\rm BZ} d {\bm k}~\widetilde{\Omega}_{\tau s}({\bm k})/{2\pi} $ has the form of the usual Chern number, and $\widetilde{\Omega}_{\tau s}({\bm k})$ is formally similar to the Berry curvature (see Appendix \ref{app:piezo_const})
\begin{align}
\widetilde{\Omega}_{\tau s} ({\bm k}) &= \frac{i}{\left [ d^{({\rm cv})}_{\tau s}({\bm k})\right]^2} 
\Big \{  \left \langle u^{({\rm v})}_{\tau s}({\bm k}) \right |  \widetilde{v}_{\tau s,x}({\bm k}) \left | u^{({\rm c})}_{\tau s}({\bm k}) \right \rangle 
\nonumber \\ &\times
\left \langle u^{({\rm c})}_{\tau s}({\bm k}) \right | v_{\tau s,y}({\bm k})  \left | u^{({\rm v})}_{\tau s}({\bm k}) \right \rangle 
 - (x \leftrightarrow y)
\Big \}~.
\end{align}
Here, $d^{({\rm cv})}_{\tau s}({\bm k})=E^{({\rm c})}_{\tau s}({\bm k})-E^{({\rm v})}_{\tau s}({\bm k}) $, and $E^{({\rm c/v})}_{\tau s}({\bm k})$ are the energy dispersion of the conduction/valence band. Notice that $v_{\tau s,\alpha}({\bm k})=\partial {\cal H}_{\tau s}({\bm k})/\partial k_\alpha$ is the standard band velocity, and the term $\widetilde{v}_{\tau s,\alpha}({\bm k})= a_0  \partial {\cal H}_{\tau s}({\bm k}, { \bm {\mathcal A}})/\partial {\cal A}_\alpha |_{\small {\cal A} \to 0} $ can be termed as ``{\it fictitious velocity}". For a generic two-band model for each (spin,valley) pair we can write ${\cal H}_{\tau s}({\bm k}) = \epsilon_{\tau s}({\bm k}) \mathbb{1} + {\bm h}_{\tau s}({\bm k}) \cdot {\bm \sigma }$, where 
${\bm h}_{\tau s} =(h_{\tau s,x} , h_{\tau s,y} , h_{\tau s,z})$ and  ${\bm \sigma}= (\sigma_x,\sigma_y,\sigma_z)$ are Pauli matrices. The two-band model Hamiltonian of the strained crystal can be expressed as
 \begin{equation}\label{Eq:Hstrain}
{\cal H}_{\tau s}({\bm k},{ \bm {\mathcal A}}) = \epsilon_{\tau s}  \left({\bm k}+ \frac{\eta_0{ \bm {\mathcal A}}}{2a_0} \right) \mathbb{1}
+\sum_{i} h_{\tau s,i} \left({\bm k}+  \frac{\eta_i{ \bm {\mathcal A}}}{2a_0}  \right)  \sigma_i~,
\end{equation}
where $i=x,y,z$ and $(\eta_0,\eta_i)$ are dimensionless parameters accounting for the strength of particle-strain coupling. 
Particularly, Eq.~(\ref{Eq:Hstrain}) has been obtained explicitly for graphene, monolayer h-BN and monolayer TMDs. After Eq.~(\ref{Eq:Hstrain}), we can evaluate the velocity as  ${\bm v}_{\tau s}({\bm k}) = {\bm \nabla} {\cal H}_{\tau s}({\bm k})$,  and the fictitious velocity as 
$\widetilde{\bm v}_{\tau s}({\bm k}) = \left \{ \eta_0 {\bm \nabla}\epsilon  ({\bm k}) \mathbb{1}+ \sum_{i} \eta_i {\bm \nabla} h_{\tau s,i} ({\bm k} )  \sigma_i \right \} /2$.
Notice that, for the simplest graphene-like case, $\eta_{i=0,x,y,z}=\eta$, the fictitious velocity is proportional to the velocity. In this case, therefore ${\Omega}_{\tau s} ({\bm k})=2\widetilde{\Omega}_{\tau s} ({\bm k})/\eta $ coincides with the conventional Berry curvature and ${\cal C}_{\tau s}=2\widetilde{\cal C}_{\tau s}/\eta $ is the Chern number. 
\par
{\it Piezoelectric constant of h-BN and TMDs.---}
In the following, we apply the developed theory to calculate the piezoelectric constant of two paradigmatic families of 2D crystals with $D_{3h}$ symmetry: h-BN and TMDs. The effective ${\bm k\cdot \bm p}$ Hamiltonian of h-BN in the ``sublattice" space is given by $\epsilon_{\tau}({\bm k})=0$ and ${\bm h}_{\tau}({\bm k}) = (\hbar v \tau k_x, \hbar v k_y, \Delta )$, where ${\bm k}=(k_x, k_y)$ is the relative momentum with respect to the K-point of the BZ, $\hbar v=3t_0 a_0/2$, where $t_0 \sim 2.3~{\rm eV}$, $\Delta\sim 5.97~{\rm eV}$, and $a=\sqrt{3} a_0=2.5~{\rm \AA}$, are the nearest neighbor hopping, band gap and lattice constant, respectively~\cite{Robertson_1984,Watanabe_2004,Droth_2016}. The spin degree of freedom leads to a double degeneracy of the states and therefore we drop the subindex $``s"$ in the h-BN Hamiltonian. 
On the other hand, the effective ${\bm k\cdot \bm p}$ model for  monolayer TMDs (ignoring trigonal warping effects) in ``band'' space is \cite{Rostami_2015}  
\begin{align}
&\epsilon_{\tau s}({\bm k})= \frac{\Delta_0+\lambda_0\tau s}{2}+\frac{\hbar^2 |{\bm k}|^2 }{4m_0 } \alpha~, 
\nonumber \\
&{\bm h}_{\tau s}({\bm k}) = \Big (t_0 a_0 \tau k_x,t_0 a_0 k_y, \frac{\Delta+\lambda\tau s}{2}+\frac{\hbar^2 |{\bm k}|^2 }{4m_0 } \beta  \Big )~.
\end{align}
Here, $m_0$, the free electron mass, $a_0$, $t_0$, $\Delta_0$, $\Delta$, $\lambda_0$, $\lambda$, $\alpha$, and $\beta$ are strain-independent parameters that can be obtained in terms of the Slater-Koster parameters of the original tight-binding Hamiltonian. Numerical values for the different monolayer TMDs considered in this work are given in Table~\ref{tab:TMD_k.p}.\vspace{3mm}
\begin{table}[h]
\caption{${\bm k}\cdot{\bm p}$ parameters of TMDs extracted from 
 the low-energy projection of a tight-binding model~\cite{Rostami_2015,Rafael_as_2016}. The lattice parameters are taken from Ref. \cite{Duerloo_2012}.}
  \begin{tabular}{  c  c  c  c  c  }
    \hline\hline
    & \hspace{4mm}MoS$_2$\hspace{4mm} & \hspace{4mm}MoSe$_2$\hspace{3mm} &\hspace{3mm}WS$_2$\hspace{4mm}&\hspace{5mm}WSe$_2$\hspace{4mm}  \\
    \hline \hline  
    $a=\sqrt{3} a_0[{\rm \AA}]$   & 3.160   & 3.290    &3.150   &3.290   \\  
    $t_0[{\rm eV}]$             &2.338    &2.110    &3.274   &2.683   \\  
    $\Delta[{\rm eV}]$        &1.823     & 1.468   &1.787  &1.576  \\  
    $\lambda[{\rm eV}]$     &-0.092   &-0.111   &-0.265  &-0.281   \\
    $\Delta_0[{\rm eV}]$    &-0.110    &-0.436 & 0.049   &-0.343   \\  
    $\lambda_0[{\rm eV}]$& 0.080   &0.067    &0.251  &0.218   \\  
    $\alpha$                       &-0.010   &-0.093   &-0.308  &0.184 \\  
    $\beta$                         &-1.540   &-1.367   &-1.914 &-1.892  \\
    $\eta_0/2$                    &-50.544 &-5.532  &-3.447  &7.073   \\   
    $\eta_x/2=\eta_y/2$    &0.002  &0.0353   &0.033&0.071   \\  
    $\eta_z/2$                    &1.635  &1.560   &1.923   &1.440  \\  
    \hline\hline
  \end{tabular}
\label{tab:TMD_k.p}
\end{table}
\par
In strained h-BN, we only have one independent particle-strain coupling, $\eta_x=\eta_y=\eta\sim3.3$ \cite{Droth_2016} leading to the simple relation $\widetilde{\bm v}_{\tau} = \eta {\bm v}_{\tau} /2$. As a consequence, 
$\widetilde{\cal C}_{\tau}$ is  proportional to the usual Chern number in the massive Dirac model, i.e. ${\cal C}_{\tau} =  \tau{\rm sign}[\Delta]/2$~, and we find
\begin{equation}
\gamma^{\text{h-BN}} =  \eta \frac{e}{4\pi a_0}   {\cal C}_{\rm valley}~,
\end{equation}
where the {\it valley Chern number} is defined by ${\cal C}_{\rm valley} = \sum_{\tau} \tau {\cal C}_{\tau}={\rm sign}[\Delta] $. 
This result differs from the one reported in Ref. \cite{Droth_2016}, which contains a high-energy cutoff that does not appear in our derivation. Therefore, measurements of the piezoelectric constant can be used as {\it direct} tools to analyze the valley Chern number. 
Topological valley currents have been recently detected through nonlocal transport measurement in multi-terminal devices~\cite{Gorbachev_2014,Shimazaki_2015,Sui_2015}. 
Here, we propose that piezoelectricity measurements can be used to access the valley degree of freedom in systems like h-BN, whose large gap impedes non-local transport experiments like those performed in graphene superlattices~\cite{Gorbachev_2014} and bilayer graphene~\cite{Shimazaki_2015,Sui_2015}. We notice that applying time-dependent strain can induce a synthetic valley-dependent electric field which can  derive charge current in topological systems such as gapped graphene~\cite{Oppen_2009,Vaezi_2013}.

The numerical value of $\gamma^{\text{ h-BN}}$ obtained from our theory is given in Table.~\ref{tab:gamma}, showing good agreement with {\it ab initio} calculations. 
\par
The case of strained TMDs is more complex, since $\eta_x=\eta_y$ while they differ from $\eta_0$ and $\eta_z$. Contrary to the simpler h-BN case, the fictitious velocity in TMDs is not proportional to the velocity and consequently we cannot use the simplified relation with the usual Chern number. However, we still can evaluate $\widetilde{\cal C}_{\tau s}$ explicitly from the TMD's Hamiltonian. After a straightforward calculation, we arrive at the following analytical expression for the piezoelectric constant of TMDs (see Appendix \ref{app:piezo_tmd})
\begin{align}\label{eq:gamma_tmd}
\gamma^{\text{TMDs}}= (\eta_x+\eta_z) \frac{  e}{8\pi a_0}  \sum_{s=\pm} \frac{{\cal D}_s(\eta_x,\eta_z)+2 \beta \Lambda_s  {\cal C}_s}{1+2\beta \Lambda_s}~, 
\end{align}
where $\Lambda_s ={\hbar^2 (\Delta+s\lambda)}/{(4m_0 t^2_0a^2_0)}$~,
\begin{align}
{\cal C}_s  & =  \frac{{\rm sign}[\Delta+s \lambda]-  {\rm sign}[\beta]}{2}~,
\end{align}
and
\begin{align}
{\cal D}_s (\eta_x,\eta_z)& =  \frac{\eta_x {\rm sign}[\Delta+s \lambda]-\eta_z  {\rm sign}[\beta]}{\eta_x+\eta_z}~.
\end{align}
Notice that ${\cal C}_{s} $ is the usual K-valley $(\tau=+)$ Chern number of monolayer TMDs with spin $s=\pm$. Intriguingly, depending on the relative sign of $\beta$ and $\Delta\pm\lambda$, we either have a topological (${\cal C}_{s}=\pm1$) or a trivial  (${\cal C}_{s}=0$) phase in each valley~\cite{Rostami_prb_2013,Rostami_jpcm_2016}. This topological property is protected as long as inter-valley scattering is suppressed. Since $\Delta\pm\lambda>0$ for the case of interest here, one can simplify Eq.~(\ref{eq:gamma_tmd}). We find 
\begin{align}\label{Eq:GammaBeta}
\gamma^{\text{TMDs}}  =\frac{e}{4\pi a_0} \begin{cases} 
      \eta_x+\eta_z
     &\text{$\beta<0 $} \\ 
        \eta_x 
     &\text{$\beta=0$} \\ 
    \sum_{s=\pm} \frac{\eta_x-\eta_z}{2+4\beta \Lambda_s}
     &\text{$\beta>0$} 
\end{cases}~.
\end{align}
The values of $\gamma^{\text{TMDs}}$ obtained from our ${\bm k}\cdot{\bm p}$ method are shown in Table \ref{tab:gamma}. Again, in spite of the simplicity of our model, the results that we find are in good agreement with existing {\it ab-initio} and experimental results, strengthening the validity of our approach and providing microscopic insight into piezoelectricity in 2D crystals. 
\begin{table}[h]
\caption{The numerical value of piezoelectric constant $\gamma~[{\rm 10^{-10} C/m}]$ obtained here (${\bm k}\cdot{\bm p}$ method), and the previously reported DFT (clamped-ion) and  experimental results.}
  \begin{tabular}{ c c  c  c  c  c  }
    \hline\hline
    &h-BN&MoS$_2$ &MoSe$_2$&WS$_2$&WSe$_2$  \\
    \hline  \hline  
   This work   &2.91        &2.29   &2.14    &2.74   &2.03  \\ 
    DFT\cite{Duerloo_2012,Michel_2017}      &3.71  &3.06     & 2.80    &2.20    &1.93     \\  
    Exp.\cite{Zhu_2015}   &---  & $2.9\pm0.5$   &---   &---  &---  \\  
  \hline\hline
  \end{tabular}
\label{tab:gamma}
\end{table}
\par
{\it Effect of inhomogeneous strain.---} In the following we consider the polarization induced in samples subjected to inhomogeneous strain. This is a highly relevant problem due to the large number of recent experiments in which  2D crystals are subjected to a non-uniform strain profile~\cite{Levy_2010,Yan_2012,Gomes_2012,Lu_2012,Khestanova_NC_2016,Castellanos-Gomez_NL_2013,Quereda_NL_2016,Li_NC_2015,Branny_2017,Polini2013,Palacios-Berraquero_2017}. Neglecting, for long-wavelength deformations, the flexoelectric (i.e. a term accounting for the polarization induced by the strain gradient~\cite{Yudin_2013}), we can generalize to the inhomogeneous strain case the linear-response relation for the piezoelectric tensor: $P_i({\bm r}) \approx \sum_{j k} \gamma_{i j k} \varepsilon_{j k}({\bm r})$. Consequently the induced charge density, following Eq.~(\ref{eq:P_sym}), reads $ \rho({\bm r})=en({\bm r}) = - {\bm \nabla}\cdot{\bm P}({\bm r}) \approx - \gamma  ~ \hat{\bm z} \cdot \left [{\bm \nabla} \times {\cal {\bm A}}({\bm r}) \right ]$. 
The dependence of $\rho ( {\bm r} )$ on the strain tensor is the same as the dependence of the strain-induced pseudomagnetic field acting on electrons in graphene. Unlike this case, the induced charge density has the same sign in the two valleys.
\par
We can obtain simple estimates of the charge density induced by a variation in strain, $\Delta \varepsilon$, over a length $\ell$. The variation of the strain leads to ${\bm \nabla} \times {\cal {\bm A}}({\bm r}) \sim \Delta \varepsilon / \ell$, so that $n ( {\bm r} ) \sim (\gamma/e) \Delta \varepsilon / \ell \sim   \Delta \varepsilon / ( a_0 \ell )$. The materials considered here can withstand large strains. In MoS$_2$ or h-BN bubbles~\cite{Khestanova_NC_2016}, the variations in the strain can be of order $\Delta \epsilon \approx 0.02$ over scales of  $\ell \sim 300$ nm, leading to $n \sim 10^{11} {\rm cm}^{-2}$. Higher strain gradients, with maximum strains $\Delta \epsilon \sim 0.1$ over short lengths, $\ell \sim 10 - 15$ nm have been reported in graphene bubbles on metallic surfaces~\cite{Levy_2010}. Similar configurations will induce carrier densities $n \sim 10^{13} {\rm cm}^{-2}$. We notice that, for larger values of applied strain, second-order piezoelectric effects might be relevant for an accurate estimation of the induced carrier density. This is the case in Zinc-Blende \cite{Bester_PRL_2006} and wurtzite \cite{Pal_PRB_2011} semiconductors.
\par
In order to illustrate further the charge induced by non-uniform strains, we discuss in detail the case of MoS$_2$ and h-BN bubbles described in \cite{Khestanova_NC_2016}. The shape of these bubbles is determined by the competition between the elastic energy of the 2D material and the van der Waals attraction to the substrate. We consider bubbles with radial symmetry. The shape and internal strains are defined by the in plane and out of plane displacements, $u( r ) , h ( r )$. The form of these functions are universal, and determined by the ratio $h_{\rm max} / R$, where $h_{\rm max}$ is the height of the bubble and $R$ is its radius. The polarization vector for this case is given by ${\bm P}({\bm r})=  p(r) [ \hat{\bm r}  \sin(3\theta)  +\hat{\bm \theta} \cos(3\theta) ]$ where (see Appendix \ref{app:piezo_bubbles})
\begin{align}\label{eq:p_tilde}
 p(r)=  \gamma\left \{ \frac{ u(r)}{r} -  \frac{\partial  u(r)}{\partial r} - \frac{1}{2} \left (\frac{\partial h (r)}{\partial r}\right ) ^2 \right \} = p_0 \widetilde{p} \left( \frac{r}{R} \right)~.  
\end{align}
Here, $p_0=\gamma{(1+\nu)h_{\rm max}^2}/{R^2}$, with $\nu$ the Poisson's ratio, and $\widetilde{p} ( x )$ is a universal function which does not depend on the material. The induced charge density is
\begin{align}
\rho ( {\bm r} ) &=  \rho_0 \widetilde{\rho}  \left( \frac{r}{R} \right) \sin ( 3 \theta )
\end{align}
where $\rho_0=p_0/R$ and, as before, the function $\widetilde{\rho} ( x )$ is universal. This analysis is consistent with our previous estimates, as $\Delta \varepsilon \sim h_{\rm max}^2 / R^2$, and $\ell \sim R$. The charge distribution is shown in Fig. \ref{fig}. The charge density reflects the trigonal symmetry of the lattice, and, as a result, it vanishes at the apex of the bubble, $r = 0$. We assume that the piezoelectric layer slides and relaxes outside the region where it is detached from the substrate. As a result, the charge density decays as $r^{-3}$ outside the bubble. Note that the aspect ratio, $h_{\rm max} / R$ is independent of the size of the bubble, so that the size dependence of the induced charge density is $R^{-1}$. For $\gamma \sim 1$ and $h_{\rm max} / R \sim 0.1$, we find $\rho \sim 10^{7} / R$ in units of electron charge $\times$ cm$^{-2}$. For $R \approx 1 \mu$m, we obtain $\rho_0 \approx 10^{11} e \times {\rm cm}^{-2}$.  For more details, see Appendix \ref{app:piezo_bubbles}. 
Finally, a number of schemes have been proposed to study gauge fields in graphene
~\cite{GGKN10,LGK11,ZSL15}. If the graphene layer in devices, such  as quantum emitters~\cite{KKG15,Branny_2017,Palacios-Berraquero_2017}, was encapsulated in h-BN, the pseudomagnetic fields in graphene and the charge density induced in the h-BN layers are proportional. For example, a pseudomagnetic field of one Tesla in graphene implies $n \approx 10^{11} {\rm cm}^{-2}$~.
\begin{figure}[h!]
\centering
\begin{overpic}[width=0.48\linewidth]{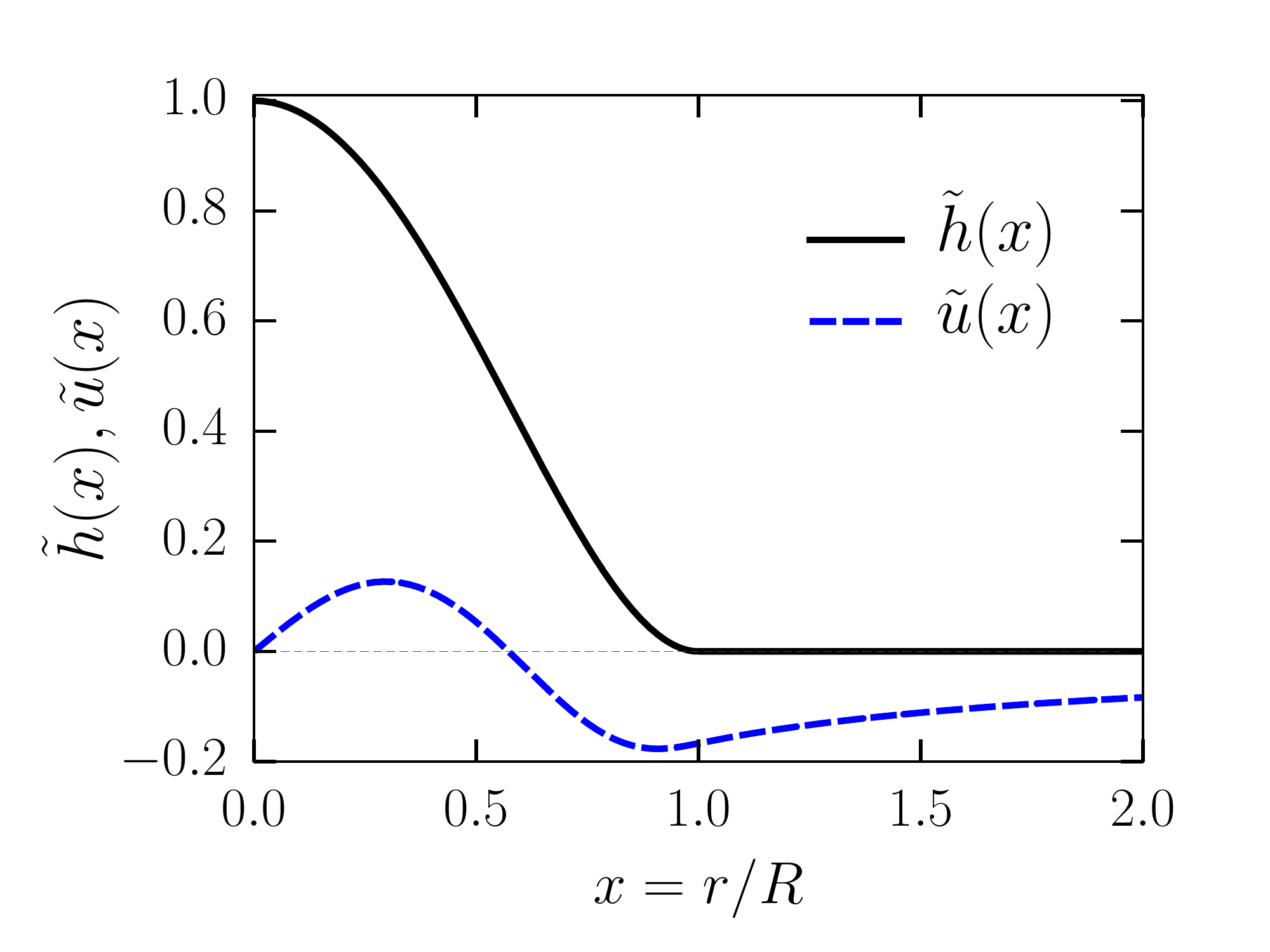}\put(1,76){(a)}\end{overpic}
\begin{overpic}[width=0.5\linewidth]{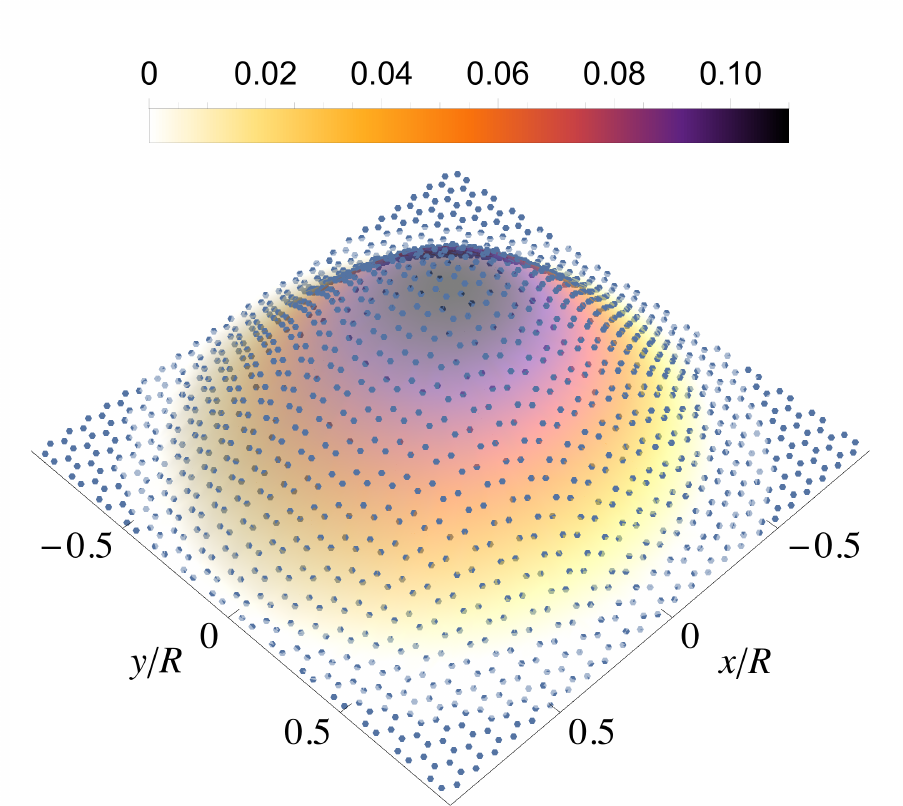}\put(1,72){(b)}\end{overpic}
\begin{overpic}[width=0.70\linewidth]{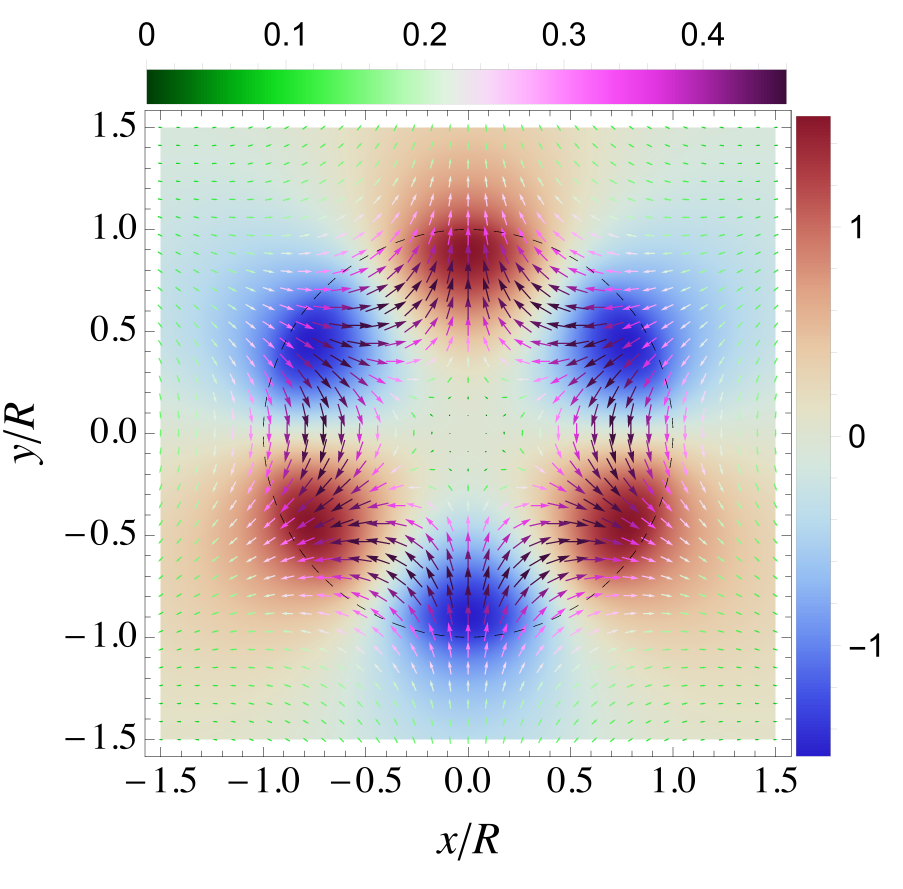}\put(1,85){(c)}\end{overpic}
\caption{ 
(a) Radial and out-of-plane displacement profiles corresponding to a pure-bending deformation with $\widetilde h(x) \equiv h(x)/h_{\rm max}= (1 - x^2)^2 \Theta(1-x)$ and the dimensionless radial displacement, $\widetilde u(x)\equiv u(x)/u_0$ with $u_0=h^2_{\rm max}/R$, obtained from elasticity theory (see Appendix \ref{app:piezo_bubbles}). The horizontal red dashed line is given as a guide to the eye, showing the zero displacement level.  
(b) Bubble shape with a pure-bending deformation profile with $h_{\rm max}\sim 0.11R$.
(c) Polarization field, $\widetilde{p}(x) [ \hat{\bm r}  \sin(3\theta)  +\hat{\bm \theta} \cos(3\theta) ]$, streamlines on top of the induced charge density, $\widetilde{\rho}(x) \sin (3 \theta )$, colormap. 
The horizontal (vertical) colorbar corresponds to the streamline (colormap) plot.
The dashed circle indicates the bubble's boundary at $r=R$.}
\label{fig}
\vspace{-2mm}
\end{figure}
\par
{\it Conclusions.---} 
In summary, we have performed a systematic study of the piezoelectric response of non-centrosymmetric hexagonal 2D crystals. Starting from a general ${\bm k} \cdot {\bm p}$ model Hamiltonian, we have obtained a closed analytical expression for the piezoelectric constant in terms of the {\it valley Chern number},  bridging valleytronics valleytronics~\cite{valleytronics2015,valleytronics2016} and piezotronics~\cite{Wang2010}. The particular cases of h-BN and TMDs (MoS$_2$, WS$_2$, MoSe$_2$ and WSe$_2$) have been studied. The validity of the theory has been proven by the good quantitative agreement found between the piezoelectric constant obtained from our method, and that calculated from {\it ab initio} approaches and experimental measurements. 
\par
We finally generalize the theory to study samples subjected to inhomogeneous strain, which is a case of great experimental interest and which cannot be studied with standard DFT methods due to the computational cost. We demonstrate that piezoelectric effect in inhomogenous crystals leads to the appearance of significant carrier densities in the sample. 

{\it Acknowledgments.---} 
This work has received funding from the European Unions Seventh Framework Programme (FP7/2007-2013) through the ERC Advanced Grant NOVGRAPHENE (GA No. 290846), European Commission under the Graphene Flagship, contract CNECTICT-604391, the Spanish MINECO through Grants No. FIS2014- 58445-JIN and RYC-2016-20663, Fondazione Istituto Italiano di Tecnologia, the European Union's Horizon 2020 research and innovation programme under grant agreement No. 696656 ``GrapheneCore1". 

\newpage
\appendix
\begin{widetext}
\section{Symmetry consideration}\label{app:sym}
Within the linear response theory, the strain-induced polarization reads
\begin{equation}\label{eq:P_epsilon}
P_{i} = \sum_{j k} \gamma_{i j k } \varepsilon_{j k}~,
\end{equation}
where ${\bm P}$ and ${\bm \varepsilon}$ are the polarization vector and strain tensor respectively, and $\gamma_{i j k}$ stands for the piezoelectric tensor element. 
For the rank-3 tensor ${\bm \gamma}$, we can obtain the following symmetry relation~\cite{Arfken}
\begin{equation}\label{eq:sym}
\sum_{i' j' k'} R^\dagger_{i i'} \gamma_{i' j' k'} R_{j' j} R_{k' k} = \gamma_{ijk}~,
\end{equation}
where ${\bm R}$ stands for the matrix representation of a symmetry operator.
Considering three-fold rotational, $C_3$, and vertical mirror, $\sigma_{\rm v}:x \to -x$, symmetries in $D_{3h}$ point group ---see Fig.~\ref{lattice}, one can conclude
 \begin{figure}[h!]
 \centering 
 \begin{overpic}[width=0.4\linewidth]{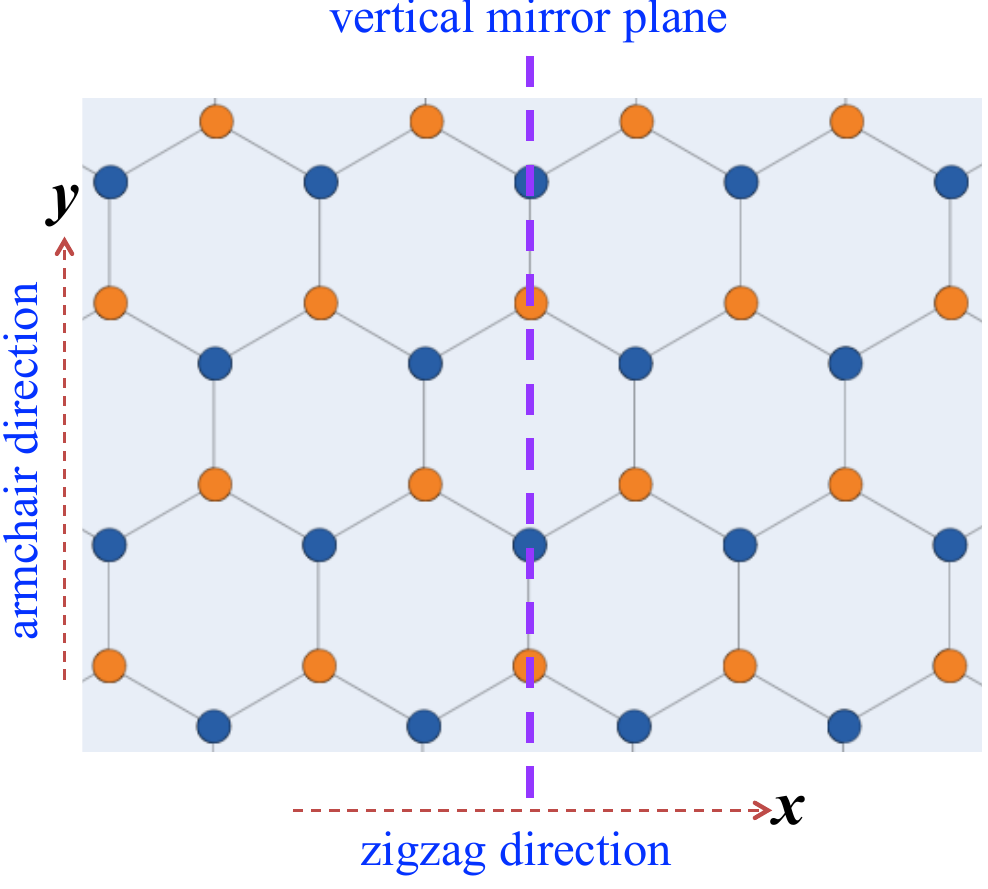}\put(5,66){}\end{overpic}
\caption{Sketch of lattice structure for $D_{3h}$ symmetric crystals such as h-BN and TMDs.  Solid circles with different color denote two inequivalent sublattices.}
\label{lattice}
\end{figure}
\begin{eqnarray}\label{eq:sym}
&&\gamma_{xxx}=\gamma_{xyy}=\gamma_{yxy}=\gamma_{yyx} =0~,
\nonumber \\ 
&&\gamma_{yyy} = -\gamma_{yxx}=-\gamma_{xyx} =-\gamma_{xxy}~.
\end{eqnarray}
Notice that 
\begin{equation}
{\bm R}(C_3) \equiv \begin{bmatrix} -\frac{1}{2}& \frac{\sqrt{3}}{2} \\ \\  - \frac{\sqrt{3}}{2} &  -\frac{1}{2} \end{bmatrix}
~,~~~~~ 
{\bm R}(\sigma_ {\rm  v}) \equiv \begin{bmatrix} -1& 0 \\ 0& 1 \end{bmatrix}~.
\end{equation}
Owing to the mirror symmetry, all tensor elements with an odd number of $x$ Cartesian index are identically zero. 
Considering  the symmetry relations given in Eq.~(\ref{eq:sym}), we can rewrite Eq.~(\ref{eq:P_epsilon}) as follows
\begin{equation}\label{eq:P_sym} 
{\bm P}= \gamma_{yyy} {\bm {\mathcal A}} \times \hat {\bm z}~,
\end{equation}
where  
\begin{equation}\label{eq:cal_A}
{\bm {\mathcal A}} =  (\varepsilon_{xx}-\varepsilon_{yy}) \hat{\bm x} - 2 \varepsilon_{xy}  \hat{\bm y}~.
\end{equation}
According to this relation the charge polarization is always perpendicular to the vector $ {\bm {\mathcal A}}$.
%

\section{Piezoelectric constant}\label{app:piezo_const} 
The electronic polarization is given by~\cite{KS_DV_1993,Resta_1994,Resta_Vanderbilt,Resta_2010}
\begin{equation}\label{eq:p_el}
{\bm P}({ \bm {\mathcal A}})  = -e \sum_{\tau s} \int_{\rm BZ} \frac{d {\bm k} }{(2\pi)^2}  { \bm a}^{({\rm v})}_{\tau s}({\bm k}, \tau { \bm {\mathcal A}})~,
\end{equation}
where  $\tau {\bm {\mathcal A}}$ is the fictitious gauge field at $\tau$-valley and the superscript (v) stands for the valence band index.  The piezoelectric constant reads 
\begin{equation}
\gamma  \equiv \gamma_{yyy}= \frac{\partial {P}_{y}({\bm {\mathcal A}})}{\partial \varepsilon_{yy}}\Big |_{{\bm {\mathcal A}} \to {\bm 0}}  
=  - \frac{\partial {P}_{y}({\bm  {\mathcal A}})}{\partial {\cal A}_{x}}\Big |_{{\bm {\mathcal A}} \to {\bm 0}} =
e \sum_{\tau s} \int_{\rm BZ} \frac{d {\bm k} }{(2\pi)^2}  \frac{\partial a^{({\rm v})}_{\tau s,y}({\bm k}, \tau { \bm {\mathcal A}})}{\partial {\cal A}_{x}}\Big |_{{\bm {\mathcal A}} \to {\bm 0}}~.
\end{equation}
We can use the following decomposition 
\begin{equation}\label{Eq:da/dAx}
\frac{\partial a^{({\rm v})}_{\tau s,y}({\bm k}, \tau { \bm {\mathcal A}})}{\partial {\cal A}_{x}}
  =
  \frac{1}{2} \left \{  
\frac{\partial a^{({\rm v})}_{\tau s,y}({\bm k}, \tau { \bm {\mathcal A}})}{\partial {\cal A}_{x}}
   +   
\frac{\partial a^{({\rm v})}_{\tau s,x}({\bm k}, \tau { \bm {\mathcal A}})}{\partial {\cal A}_{y}} \right \} 
   +
     \frac{1}{2} \left \{   
\frac{\partial a^{({\rm v})}_{\tau s,y}({\bm k}, \tau { \bm {\mathcal A}})}{\partial {\cal A}_{x}}
   -   
\frac{\partial a^{({\rm v})}_{\tau s,x}({\bm k}, \tau { \bm {\mathcal A}})}{\partial {\cal A}_{y}} \right \}~.
\end{equation}
From Eq.~(\ref{eq:P_sym}) we have that $ {\partial {P}_{y}({\bm  {\mathcal A}})}/{\partial {\cal A}_{x}} = -  {\partial {P}_{x}({\bm  {\mathcal A}})}/{\partial {\cal A}_{y}} $. On the other hand, the first  term in Eq.~(\ref{Eq:da/dAx}) does not respect such anti-symmetric relation under $x \leftrightarrow y$ exchange.  
Therefore, the integral of the first symmetric term in the above relation on whole BZ is zero based on the D$_{3h}$ symmetry. 
In this regard, we reach the following relation which is  identical to Eq.~(3) of the main text. 
\begin{equation}
\gamma = 
\frac{e}{2} \sum_{\tau s} \int_{\rm BZ} \frac{d {\bm k} }{(2\pi)^2} 
\left \{   
\frac{\partial a^{({\rm v})}_{\tau s,y}({\bm k}, \tau { \bm {\mathcal A}})}{\partial {\cal A}_{x}}
   -   
\frac{\partial a^{({\rm v})}_{\tau s,x}({\bm k}, \tau { \bm {\mathcal A}})}{\partial {\cal A}_{y}} 
\right \}_{{\bm {\mathcal A}} \to {\bm 0}} 
\end{equation}
Using the definition of Berry connection, we can obtain 
\begin{eqnarray}
\frac{\partial a^{({\rm v})}_{\tau s,y}({\bm k}, \tau { \bm {\mathcal A}})}{\partial {\cal A}_{x}}
   -   
\frac{\partial a^{({\rm v})}_{\tau s,x}({\bm k}, \tau { \bm {\mathcal A}})}{\partial {\cal A}_{y}} 
   =  
i\left \{   
 \left \langle \frac{\partial u^{({\rm v})}_{\tau s}({\bm k},\tau {\bm {\mathcal A}})}{\partial {\cal A}_{x}} \bigg |  \frac{\partial u^{({\rm v})}_{\tau s}({\bm k},\tau {\bm {\mathcal A}})}{\partial k_{y}}  \right \rangle 
   -
 \left \langle \frac{\partial u^{({\rm v})}_{\tau s}({\bm k},\tau {\bm {\mathcal A}})}{\partial {\cal A}_{y}} \bigg |  \frac{\partial u^{({\rm v})}_{\tau s}({\bm k},\tau {\bm {\mathcal A}})}{\partial k_{x}}  \right \rangle 
      \right \}~.
\end{eqnarray}
We linearize the strain-electron interaction part: 
\begin{equation}
{\cal H}_{\tau s}({\bm k},\tau{ \bm {\mathcal A}}) \approx {\cal H}_{\tau s}({\bm k}) 
+ \frac{\tau}{a_0} \sum_{\alpha}  {\cal  A}_\alpha  \widetilde{v}_{\tau s,\alpha}({\bm k}) + {\cal O}({\cal A}^2)~.
\end{equation} 
Notice that $a_0$ is an effective lattice constant and the fictitious velocity is defined as follows 
\begin{equation}
 \widetilde{\bm v}_{\tau s}({\bm k}) = a_0  \frac{\partial {\cal H}_{\tau s}({\bm k}, { \bm {\mathcal A}})}{\partial {\cal A}_\alpha} \Big |_{{\bm {\mathcal A}} \to {\bm 0}}~.
\end{equation}
Within first order perturbation theory, the Bloch eigen-function reads 
\begin{equation}
\left | u^{(n)}_{\tau s}({\bm k},\tau {\bm {\mathcal A}}) \right \rangle =   \left | u^{(n)}_{\tau s}({\bm k}) \right \rangle 
+\frac{ \tau}{a_0} \sum_{m \neq n} 
\frac{\left | u^{(m)}_{\tau s}({\bm k}) \right \rangle }{E^{(n)}_{\tau s}({\bm k})-E^{(m)}_{\tau s}({\bm k})} 
 \left \langle u^{(m)}_{\tau s}({\bm k}) \right | \widetilde{v}_{\tau s}({\bm k})\cdot {\bm {\mathcal A}} \left | u^{(n)}_{\tau s}({\bm k}) \right \rangle+ {\cal O}({\mathcal A}^2)~,
\end{equation} 
where the sum runs over band indices $m$. For a two-bands model, we have
\begin{eqnarray}
\left \{   
\frac{\partial a^{({\rm v})}_{\tau s,y}({\bm k}, \tau { \bm {\mathcal A}})}{\partial {\cal A}_{x}}
   -   
\frac{\partial a^{({\rm v})}_{\tau s,x}({\bm k}, \tau { \bm {\mathcal A}})}{\partial {\cal A}_{y}} 
\right \}_{{\bm {\mathcal A}} \to {\bm 0}} 
 &=&  
   i \frac{\tau}{a_0} 
\frac{\left \langle u^{({\rm c})}_{\tau s}({\bm k}) \bigg |  \frac{\partial u^{({\rm v})}_{\tau s}({\bm k})}{\partial k_{y}}  \right \rangle  }{E^{({\rm v})}_{\tau s}({\bm k})-E^{({\rm c})}_{\tau s}({\bm k})} 
 \left \langle u^{({\rm v})}_{\tau s}({\bm k}) \right | \widetilde{v}_{\tau s,x}({\bm k}) \left | u^{({\rm c})}_{\tau s}({\bm k}) \right \rangle 
\nonumber \\ &-&
   i \frac{\tau}{a_0} 
\frac{\left \langle u^{({\rm c})}_{\tau s}({\bm k}) \bigg |  \frac{\partial u^{({\rm v})}_{\tau s}({\bm k})}{\partial k_{x}}  \right \rangle  } {E^{({\rm v})}_{\tau s}({\bm k})-E^{({\rm c})}_{\tau s}({\bm k})} 
 \left \langle u^{({\rm v})}_{\tau s}({\bm k}) \right | \widetilde{v}_{\tau s,y}({\bm k}) \left | u^{({\rm c})}_{\tau s}({\bm k}) \right \rangle~. 
\end{eqnarray}
We use the following identity~\cite{Xiao}
\begin{equation}\label{eq:identity}
\left \langle u^{({\rm c})}_{\tau s}({\bm k}) \bigg |  \frac{\partial u^{({\rm v})}_{\tau s}({\bm k})}{\partial k_{\alpha}}  \right \rangle=
\frac{ 
 \left \langle u^{({\rm c})}_{\tau s}({\bm k}) \right |  v_{\tau s,\alpha}({\bm k}) \left | u^{({\rm v})}_{\tau s}({\bm k}) \right \rangle 
} {E^{({\rm v})}_{\tau s}({\bm k})-E^{({\rm c})}_{\tau s}({\bm k})}~,
\end{equation}
where the velocity operator is 
\begin{equation}
v_{\tau s,\alpha}({\bm k})=\frac{\partial {\cal H}_{\tau s}({\bm k})}{\partial k_{\alpha}}~.
\end{equation}
 Therefore, we reach  
 \begin{align}
\left \{   
\frac{\partial a^{({\rm v})}_{\tau s,y}({\bm k}, \tau { \bm {\mathcal A}})}{\partial {\cal A}_{x}}
   -   
\frac{\partial a^{({\rm v})}_{\tau s,x}({\bm k}, \tau { \bm {\mathcal A}})}{\partial {\cal A}_{y}} 
\right \}_{{\bm {\mathcal A}} \to {\bm 0}} 
 &= 
   i \frac{\tau}{a_0}  
\frac{
 \left \langle u^{({\rm c})}_{\tau s}({\bm k}) \right |  v_{\tau s,y}({\bm k}) \left | u^{({\rm v})}_{\tau s}({\bm k}) \right \rangle 
 \left \langle u^{({\rm v})}_{\tau s}({\bm k}) \right | \widetilde{v}_{\tau s,x}({\bm k}) \left | u^{({\rm c})}_{\tau s}({\bm k}) \right \rangle 
 -
(x \leftrightarrow y)
 }{\left [E^{({\rm v})}_{\tau s}({\bm k})-E^{({\rm c})}_{\tau s}({\bm k}) \right]^2}~.  
\end{align}
From this we obtain the following relation for the piezoelectric constant. 
 \begin{equation}\label{Eq:gamma-chern}
\gamma = \frac{e}{4\pi a_0} \sum_{\tau s} \tau \widetilde{\cal C}_{\tau s}~,
\end{equation}
in which  
\begin{align}\label{eq:chern}
 \widetilde{\cal C}_{\tau s} = \frac{1}{2\pi}\int_{\rm BZ} d {\bm k}~\widetilde{\Omega}_{\tau s}({\bm k})~,
 \end{align}
with
\begin{align}\label{eq:Berry_Curvature}
\widetilde{\Omega}_{\tau s} ({\bm k})  =  i  
\frac{
 \left \langle u^{({\rm v})}_{\tau s}({\bm k}) \right | \widetilde{v}_{\tau s,x}({\bm k}) \left | u^{({\rm c})}_{\tau s}({\bm k}) \right \rangle 
 \left \langle u^{({\rm c})}_{\tau s}({\bm k}) \right |  v_{\tau s,y}({\bm k}) \left | u^{({\rm v})}_{\tau s}({\bm k}) \right \rangle 
 -
(x \leftrightarrow y)
 }{\left [E^{({\rm v})}_{\tau s}({\bm k})-E^{({\rm c})}_{\tau s}({\bm k}) \right]^2}~.  
\end{align}
\section{Piezoelectric constant of TMDs}\label{app:piezo_tmd} 
The two-band model Hamiltonian of strained monolayer TMDs can be formally written as follows~\cite{Rostami_2015}
 \begin{equation} 
{\cal H}_{\tau s}({\bm k},{ \bm {\mathcal A}}) = \epsilon_{\tau s}  \left({\bm k}+ \frac{\eta_0{ \bm {\mathcal A}}}{2a_0} \right) \mathbb{1}
+\sum_{i} h_{\tau s,i} \left({\bm k}+  \frac{\eta_i{ \bm {\mathcal A}}}{2a_0}  \right)  \sigma_i~,
\end{equation}
where $i=x,y,z$ and $(\eta_0,\eta_i)$ are dimensionless parameters accounting for the strength of particle-strain coupling of each term. 
Notice that $\tau/s$ indicate of the valley/spin index. 
The eigenvalues of  unstrained Hamiltonian, ${\cal H}_{\tau s}({\bm k},{ \bm0})$, read 
\begin{align}
E^{({\rm c/v})}_{\tau s}({\bm k}) = \epsilon_{\tau s}({\bm k})  \pm | {\bm h}_{\tau s}({\bm k})|~,
\end{align}
where
\begin{equation} 
| {\bm h}_{\tau s}({\bm k})|=\sqrt{h_{\tau s,x}({\bm k})^2+ h_{\tau s,y}({\bm k})^2+h_{\tau s,z}({\bm k})^2}~.
\end{equation}
The corresponding eigenvectors are  
\begin{align}
\left |u^{({\rm c/v})}_{\tau s}({\bm k})  \right \rangle = \frac{1}{\sqrt{h_{\tau s,x}({\bm k})^2+ h_{\tau s,y}({\bm k})^2 +  \left [ \pm  | {\bm h}_{\tau s}({\bm k})|-h_{\tau s,z}({\bm k}) \right]^2}} \begin{pmatrix} h_{\tau s,x}({\bm k})- i  h_{\tau s,y}({\bm k}) \\ \\ \pm  | {\bm h}_{\tau s}({\bm k})|-h_{\tau s,z}({\bm k}) \end{pmatrix}~.
\end{align}
Note that $+/-$ sign corresponds to the conduction(c)/valence(v) band.  According to the definition of velocity and fictitious velocity, we have 
\begin{align}
&{\bm v}_{\tau s}({\bm k}) =  \frac{\partial \epsilon_{\tau s}({\bm k}) }{\partial {\bm k}}  \mathbb{1} 
+ \sum_{i} \frac{\partial  h_{\tau s, i}({\bm k}) }{\partial {\bm k}}  \sigma_i~,
\\
&\widetilde{\bm v}_{\tau s}({\bm k}) = \frac{1}{2} \left \{  \eta_0  \frac{\partial \epsilon_{\tau s}({\bm k}) }{\partial {\bm k}}  \mathbb{1} 
+ \sum_{i} \eta_i \frac{\partial  h_{\tau s, i}({\bm k}) }{\partial {\bm k}}  \sigma_i \right 
\}~, 
\end{align}
and their matrix elements are given by 
 \begin{align}
 &\left \langle u^{({\rm c})}_{\tau s}({\bm k}) \right |  v_{\tau s,y}({\bm k}) \left | u^{({\rm v})}_{\tau s}({\bm k}) \right \rangle  
 = \sum_{i}  \frac{\partial  h_{\tau s, i}({\bm k}) }{\partial  k_y } \left \langle u^{({\rm c})}_{\tau s}({\bm k}) \right |   \sigma_i  \left | u^{({\rm v})}_{\tau s}({\bm k}) \right \rangle~,
\\
 &\left \langle u^{({\rm v})}_{\tau s}({\bm k}) \right |  \widetilde{v}_{\tau s,x}({\bm k}) \left | u^{({\rm c})}_{\tau s}({\bm k}) \right \rangle  
 = \frac{1}{2} \sum_{i}  \eta_i \frac{\partial  h_{\tau s, i}({\bm k}) }{\partial  k_x } \left \langle u^{({\rm v})}_{\tau s}({\bm k}) \right |   \sigma_i  \left | u^{({\rm c})}_{\tau s}({\bm k}) \right \rangle~. 
\end{align}
Therefore, we can explicitly show that 
\begin{align}\label{eq:id1}
& 
 \left \langle u^{({\rm v})}_{\tau s}({\bm k}) \right |  \widetilde{v}_{\tau s,x}({\bm k}) \left | u^{({\rm c})}_{\tau s}({\bm k}) \right \rangle  
  \left \langle u^{({\rm c})}_{\tau s}({\bm k}) \right |  v_{\tau s,y}({\bm k}) \left | u^{({\rm v})}_{\tau s}({\bm k}) \right \rangle  
  -
 \left \langle u^{({\rm v})}_{\tau s}({\bm k}) \right |  \widetilde{v}_{\tau s,y}({\bm k}) \left | u^{({\rm c})}_{\tau s}({\bm k}) \right \rangle  
  \left \langle u^{({\rm c})}_{\tau s}({\bm k}) \right |  v_{\tau s,x}({\bm k}) \left | u^{({\rm v})}_{\tau s}({\bm k}) \right \rangle  
  =
 \nonumber \\
 &
  \frac{1}{2} \sum_{i j} 
   \eta_i 
  \left \{  
  \frac{\partial  h_{\tau s, i}({\bm k}) }{\partial  k_x }  \frac{\partial  h_{\tau s, j}({\bm k}) }{\partial  k_y }   
  -
    \frac{\partial  h_{\tau s, i}({\bm k}) }{\partial  k_y }  \frac{\partial  h_{\tau s, j}({\bm k}) }{\partial  k_x }   
  \right \} 
  \left \langle u^{({\rm v})}_{\tau s}({\bm k}) \right |   \sigma_i  \left | u^{({\rm c})}_{\tau s}({\bm k}) \right \rangle  
\left \langle u^{({\rm c})}_{\tau s}({\bm k}) \right |   \sigma_j \left | u^{({\rm v})}_{\tau s}({\bm k}) \right \rangle  
\nonumber \\
& =
  \frac{1}{2} 
 \left \{  
  \frac{\partial  h_{\tau s, y}({\bm k}) }{\partial  k_x }  \frac{\partial  h_{\tau s, x}({\bm k}) }{\partial  k_y }   
  -
    \frac{\partial  h_{\tau s, y}({\bm k}) }{\partial  k_y }  \frac{\partial  h_{\tau s, x}({\bm k}) }{\partial  k_x }   
  \right \} 
  \left \{ 
   (\eta_x-\eta_y) \frac{h_{\tau s,x}({\bm k}) h_{\tau s,y}({\bm k}) }{ |{\bm h}_{\tau s}({\bm k}) |^2 }
+ 
i (\eta_x+\eta_y) \frac{h_{\tau s,z}({\bm k}) }{|{\bm h}_{\tau s}({\bm k}) |}
  \right \}
\nonumber \\
&+
  \frac{1}{2} 
 \left \{  
  \frac{\partial  h_{\tau s, z}({\bm k}) }{\partial  k_x }  \frac{\partial  h_{\tau s, x}({\bm k}) }{\partial  k_y }   
  -
    \frac{\partial  h_{\tau s, z}({\bm k}) }{\partial  k_y }  \frac{\partial  h_{\tau s, x}({\bm k}) }{\partial  k_x }   
  \right \} 
  \left \{ 
  (\eta_x-\eta_z) \frac{h_{\tau s,x}({\bm k}) h_{\tau s,z}({\bm k})}{ |{\bm h}_{\tau s}({\bm k})|^2 }
- 
i (\eta_x+\eta_z) \frac{h_{\tau s,y}({\bm k})}{|{\bm h}_{\tau s}({\bm k})|}
  \right \}
\nonumber \\
&+
  \frac{1}{2} 
 \left \{  
  \frac{\partial  h_{\tau s, z}({\bm k}) }{\partial  k_x }  \frac{\partial  h_{\tau s, y}({\bm k}) }{\partial  k_y }   
  -
    \frac{\partial  h_{\tau s, z}({\bm k}) }{\partial  k_y }  \frac{\partial  h_{\tau s, y}({\bm k}) }{\partial  k_x }   
  \right \} 
  \left \{ 
   (\eta_y-\eta_z) \frac{h_{\tau s,y}({\bm k}) h_{\tau s,z}({\bm k})}{ |{\bm h}_{\tau s}({\bm k})|^2 }
+ 
i (\eta_y+\eta_z) \frac{h_{\tau s,x}({\bm k})}{|{\bm h}_{\tau s}({\bm k})|}
  \right \}~.
\end{align}
For the case of TMDs, we have the following explicit ${\bm k}\cdot{\bm p}$ Hamiltonian~\cite{Rostami_2015}
\begin{eqnarray}\label{eq:h_tmd}
&&\epsilon_{\tau s}({\bm k})= \frac{\Delta_0+\lambda_0\tau s}{2}+\frac{\hbar^2\alpha}{4m_0} (k^2_x+k^2_y)~, 
\nonumber \\
&&h_{\tau s,x}({\bm k}) = t_0 a_0 \tau k_x~,
\nonumber \\
&&h_{\tau s,y}({\bm k}) = t_0 a_0 k_y~,
\nonumber\\
&&h_{\tau s,z}({\bm k}) = \frac{\Delta+\lambda \tau s}{2} + \frac{\hbar^2\beta}{4m_0} (k^2_x+k^2_y)~.
\end{eqnarray}
By considering Eqs.~(\ref{eq:h_tmd}), (\ref{eq:id1}) and (\ref{eq:Berry_Curvature}), it can be seen that ${\rm Im}[\widetilde{\Omega}_{\tau s}({\bm k})] \propto k_x k_y$ and therefore its integral over whole momentum space vanishes. However, its real part is an even function of $k_x$ and $k_y$:   
\begin{align}
{\rm Re}[\widetilde{\Omega}_{\tau s} ({\bm k})]  = \tau \frac{(t_0 a_0)^2 \left \{ (\eta_x+\eta_y)\left [\frac{\Delta+\lambda \tau s}{2} + \frac{\hbar^2\beta}{4m_0} k^2\right] 
-
\frac{\hbar^2 \beta}{2m_0} [(\eta_x+\eta_z)k^2_y+(\eta_y+\eta_z)k^2_x]\right \} }{8\left ( \left[ \frac{\Delta+\lambda \tau s}{2} + \frac{\hbar^2\beta}{4m_0} k^2\right]^2 + (t_0 a_0)^2 k^2 \right )^{\frac{3}{2}}}~,
\end{align}
By using Eq.~(\ref{eq:chern}) and after performing the integral over azimuthal angle, we reach  
\begin{align}
 \widetilde{\cal C}_{\tau s} = \tau   \int^{\infty}_0 k d k   \frac{(t_0 a_0)^2 \left \{ (\eta_x+\eta_y) \frac{\Delta+\lambda \tau s}{2}  
-
\frac{\hbar^2 \beta}{2m_0} k^2 \eta_z \right \} }{8\left ( \left[ \frac{\Delta+\lambda \tau s}{2} + \frac{\hbar^2\beta}{4m_0} k^2\right]^2 + (t_0 a_0)^2 k^2 \right )^{\frac{3}{2}}}~.
 \end{align}
 In single-layer TMDs, we have $\eta_x=\eta_y$, and therefore we can simply obtain Eq.~(9) of the main text. 
\section{Piezoelectric-induced carrier density in circular bumps}\label{app:piezo_bubbles} 
We consider a 2D non-centrosymmetric crystal subjected to circularly symmetric deformation. After considering the circular symmetry, the displacement can be modeled, using polar coordinates, as $\{u_x,u_y,u_z \}= \{u(r)\cos(\theta), u(r)\sin(\theta),h(r)\}$.  
The strain tensor thus given by~\cite{Pitaevskii_Landau_Lifshitz,Timoshenko,Vozmediano}
\begin{eqnarray}\label{epsilon}
\varepsilon_{rr}(r) &=& \hat{\bm r}\cdot {\bm \varepsilon}\cdot\hat{\bm r}=\frac{\partial u(r)}{\partial r}+\frac{1}{2} \left(\frac{\partial h(r)}{\partial r} \right )^2~, 
\nonumber \\
\varepsilon_{\theta\theta}(r)&=&\hat{\bm \theta}\cdot {\bm \varepsilon}\cdot\hat{\bm \theta}=\frac{u(r)}{r}~.
\end{eqnarray}
Notice that the off-diagonal strain tensor elements are zero due to symmetry, i.e. $\varepsilon_{r \theta}= \varepsilon_{\theta r}=0$~\cite{Pitaevskii_Landau_Lifshitz,Timoshenko}.  
Using the polar coordinates, ${\bm r}=(r,\theta)$, it is straightforward to find:
\begin{eqnarray}\label{Avec}
{\cal A}_r({\bm r}) &=& \left [\varepsilon_{rr}(r)-\varepsilon_{\theta\theta}(r) \right ] \cos(3\theta)~, \nonumber\\
{\cal A}_\theta({\bm r}) &=&-\left [ \varepsilon_{rr}(r)-\varepsilon_{\theta\theta}(r) \right ] \sin(3\theta)~.
\end{eqnarray}
The electronic polarization is given by 
\begin{equation}\label{eq:polarization}
{\bm P}({\bm r}) \approx \gamma {\bm {\mathcal A}}({\bm r}) \times \hat{\bm z} =  \gamma  \left [\varepsilon_{\theta\theta}(r)-\varepsilon_{rr}(r) \right ] \left [\hat{\bm r}  \sin(3\theta)  + \hat{\bm \theta}   \cos(3\theta) \right ]~.
\end{equation}

Therefore, the induced charge density  reads
\begin{equation}\label{eq:pseudoB}
\rho({\bm r}) =-{\bm \nabla}\cdot {\bm P}({\bm r})=  \gamma (\frac{2}{r}-\frac{\partial}{\partial r}) \left[ \varepsilon_{\theta\theta}(r)-\varepsilon_{rr}(r) \right] \sin(3\theta)~.
\end{equation}
In order to calculate the polarization and induced charge density, we first need to evaluate the strain tensor. For this purpose, we define force tensor per area through the stress tensor, $\Sigma_{\alpha\beta}$, and it can be described by the Hooke's law~\cite{Pitaevskii_Landau_Lifshitz,Timoshenko}
\begin{eqnarray}\label{Hook_laws}
\Sigma_{rr}(r)   &=&              \frac{E}{1-\nu^2}         \left [ \varepsilon_{rr}(r)+\nu \varepsilon_{\theta\theta}(r) \right ]~,\nonumber \\
\Sigma_{\theta\theta}(r) &=& \frac{E}{1-\nu^2}        \left [ \varepsilon_{\theta\theta}(r)+\nu \varepsilon_{rr}(r) \right ]~. 
\end{eqnarray}
where $E$ and $\nu$ are the Young's modulus and Poisson's ratio, respectively.
At this stage, we
need to obtain $\Sigma_{rr(\theta\theta)}$.  In the absence of external in-plane force in the system, we can write the following equilibrium equation \cite{Pitaevskii_Landau_Lifshitz,Timoshenko}:
\begin{equation}
 {\bm \nabla} \cdot {\bm\Sigma} =  {\bm 0}
\end{equation}
where ${\bm\Sigma}=\hat {\bm r}  \Sigma_{rr}\hat {\bm r} + \hat {\bm \theta}  \Sigma_{\theta\theta}\hat {\bm \theta}$ is the stress tensor.
Considering $\partial \hat {\bm r}/\partial\theta =\hat {\bm \theta}$ and $\partial \hat {\bm \theta}/\partial\theta=-\hat {\bm r}$, we reach 
\begin{equation}\label{eq:sig1}
\frac{\partial \Sigma_{rr}(r)}{\partial r}+\frac{1}{r}\left [ \Sigma_{rr}(r)-\Sigma_{\theta\theta}(r) \right  ]=0~.
\end{equation}
Using Hook's laws, Eq.~(\ref{Hook_laws}), we reach
\begin{equation}\label{eq:sig1}
 \left [ \frac{\partial \varepsilon_{rr}(r)}{\partial r} +\nu  \frac{\partial\varepsilon_{\theta\theta}(r)}{\partial r}  \right  ]+\frac{1-\nu}{r}\left [ \varepsilon_{rr}(r)-\varepsilon_{\theta\theta}(r) \right  ]=0~.
\end{equation}
Using Eq.~(\ref{epsilon}), we find the following differential equation that relates the radial  and out-of-plane displacements to each other.  
\begin{equation}\label{eq:u_diff}
\frac{\partial^2 u(r)}{\partial r^2} + \frac{1}{r}\frac{\partial u(r)}{\partial r} -\frac{u(r)}{r^2} 
+ \frac{\partial h(r)}{\partial r} \frac{\partial^2 h(r)}{\partial r^2} +  \frac{1-\nu}{2 r} \left ( \frac{\partial h(r)}{\partial r} \right)^2 =0~.
\end{equation}
We define the following Green's function equation
\begin{equation}
\frac{\partial^2 g(r-r')}{\partial r^2} + \frac{1}{r}\frac{\partial g(r-r')}{\partial r} -\frac{g(r-r')}{r^2}  = -\frac{1}{r} \delta(r-r')~.
\end{equation}
Therefore, we have
\begin{equation}\label{eq:u_g_S_r}
u(r)= u_g(r)+ \int^\infty_0  g(r-r')  S(r') r' dr'~,
\end{equation}
where the source function follows 
\begin{equation}
S(r) =  \frac{\partial h(r)}{\partial r} \frac{\partial^2 h(r)}{\partial r^2} +  \frac{1-\nu}{2 r} \left ( \frac{\partial h(r)}{\partial r} \right)^2 = \frac{1}{2} \left [ \frac{\partial}{\partial r}+ \frac{1-\nu}{r}\right ] \left ( \frac{\partial h(r)}{\partial r} \right)^2~,
\end{equation}
and the general solution is 
\begin{equation}
u_g(r)= 
\left\{
	\begin{array}{ll}
		c_1 r  & \mbox{if } r \le R \\[5pt]
		c_2 r^{-1} & \mbox{if } r >R 
	\end{array}
\right.~,
\end{equation}
which contains  two unknown parameters $c_1$ and $c_2$ which can be obtained by considering continuity of the strain tensor. 
It is easy to prove that 
\begin{equation}
g(r-r')=\frac{1}{2} \frac{r_<}{r_>}~,
\end{equation}
where $r_>={\rm max}(r, r')$ and $r_<={\rm min}(r,r')$.
By plugging the above Green's function in Eq.~(\ref{eq:u_g_S_r}), we find the following solution for the radial displacement 
\begin{equation}
u(r)= u_g(r)+ \frac{1}{2} \left \{ \frac{1}{r} \int^r_0 r'^2 S(r') dr' + r  \int^\infty_r  S(r') dr'  \right \}~.
\end{equation}
Using the dimensionless variables given in Table \ref{tab:dimensionless}, we can write
\begin{table}[t]
\caption{All relevant physical quantities are given in terms of dimensionless variables and their physical unit explicitly.}
  \begin{tabular}{ c c }
		\hline\hline
	        Quantity   &  Unit \\[5pt]
	        \hline\hline
		 $r = R x$  &  $R$ \\
		$h(r) = h_{\rm max}\widetilde{h}(x)$   & $h_{\rm max}$ \\[5pt]
		$u(r) = u_0 \widetilde{u}(x)$ &   $u_0= h^2_{\rm max}/R$  \\[5pt]
	       $\{ \varepsilon_{rr}(r),\varepsilon_{\theta\theta}(r) \} = \varepsilon_0  \{ \widetilde{\varepsilon}_{rr}(x),\widetilde{\varepsilon}_{\theta\theta}(x) \} $ & $\varepsilon_0=  h^2_{\rm max}/R^2 $\\[5pt] 
	       $S(r)=S_0 \widetilde{S}(x)$ & $S_0 = h^2_{\rm max}/R^3$   \\[5pt] 
	       ${\bm P}({\bm r})= p_0 \widetilde{p}(x)    [\hat{\bm r}  \sin(3\theta)  + \hat{\bm \theta}   \cos(3\theta)   ]$ & $p_0 =\gamma (1+\nu)  {h^2_{\rm max}}/{R^2}$  \\[5pt] 
	      $ \rho({\bm r})=\rho_0 \widetilde{\rho}(x) \sin(3\theta) $ & $\rho_0 = \gamma (1+\nu)  {h^2_{\rm max}}/{R^3}$ \\[5pt] 
	       \hline\hline
  \end{tabular}
\label{tab:dimensionless}
\end{table}
\begin{equation}
\widetilde u(x)= \widetilde u_g(x)+ \frac{1}{2} \left \{ \frac{1}{x} \int^x_0 x'^2 \widetilde S(x') dx' + x  \int^\infty_x  \widetilde S(x') dx'  \right \}~,
\end{equation}
where  
\begin{equation}
\widetilde {S}(x)  = \frac{1}{2} \left [ \frac{\partial}{\partial x}+ \frac{1-\nu}{x}\right ] \left ( \frac{\partial \widetilde h(x)}{\partial x} \right)^2~.
\end{equation}
From the knowledge of the radial displacement, we can calculate the strain tensor elements by using Eq.~(\ref{epsilon}). 
Eventually, the radial profile of polarization and induced charge density can be evaluated from the following relations
\begin{equation}\label{eq:p_tilde}
\widetilde p(x)= \frac{\widetilde\varepsilon_{\theta\theta}(x) - \widetilde\varepsilon_{rr}(x)}{1+\nu} = 
\frac{1}{1+\nu } \left \{ \frac{\widetilde u(x)}{x} -  \frac{\partial \widetilde u(x)}{\partial x} - \frac{1}{2} \left (\frac{\partial \widetilde h(x)}{\partial x}\right ) ^2 \right \}~,
\end{equation}
and
\begin{equation}\label{eq:b_tilde}
\widetilde \rho(x) = \left ( \frac{2}{x} - \frac{\partial }{\partial x}  \right ) \widetilde p(x)~.
\end{equation}
The rest of this section is devoted to three particular cases of experimental interest, namely bubbles with different shapes as Gaussian, parabolic-like \cite{Khestanova_NC_2016} and pure-bending profiles. 
The parabolic-like  bubble has been discussed in Ref.~\cite{Khestanova_NC_2016}. Here, we briefly explain the  pure-bending  case for which  the bending contribution to the mechanical free energy, $\propto \left [\nabla^2 h(r)\right]^2$,  dominates the elastic energy, $\propto \left |{\bm \nabla} h(r)\right|^2$. In this approximation, the bubble profile is given by 
$D {\nabla^2} {\nabla^2} h(r) = f_z(r)$ where $D=E d^3/12(1-\nu^2)$ is the flexural rigidity.
Considering a uniform force (per area) profile as $f_z(r)=f\Theta(R-r)$, with $\Theta(x)$ the Heaviside function, and implementing $h(R)=\partial h(r)/\partial r |_{r \to R} =0$ as the boundary condition, the out-of-plane displacement reads $h(r)=h_{\rm max} \widetilde h(x)$ with  $\widetilde h(x)=\Theta(1-x)\left ( 1-x^2 \right )^2 $ and $h_{\rm max}={f R^4}/{64D}$ \cite{Pitaevskii_Landau_Lifshitz,Timoshenko}.

Straightforward calculations lead to explicit expressions for the displacement vector, $\{ \widetilde u(x),\widetilde h(x)\}$, the strain tensor, $\{\widetilde \varepsilon_{rr}(x),\widetilde \varepsilon_{\theta\theta}(x)\}$, the radial profile of electronic polarization, $\widetilde p(x)$, and the radial profile of induced charge density, $\widetilde \rho(x)$, for each of the three bubble profiles discussed in the text, as shown in Tables \ref{tab:Gaussian}, \ref{tab:parabolic} and \ref{tab:bending}. All of these mathematical functions are shown in Fig.~\ref{fig}.
\begin{table}[b]
\caption{Functional expressions for the displacement vector, strain tensor, radial profile of electronic polarization and radial profile of induced charge density, for a 2D crystal deformed following a Gaussian bubble profile.}
  \begin{tabular}{ c c }
		\hline\hline
	        Quantity   &  Functional expression \\
	        \hline\hline
	        \\[-8pt]
		$\widetilde h(x)$  &  $\exp(-x^2/2)$ \\
		$\widetilde u(x)$   & $ \frac{1}{4} x \exp(-x^2)+ \frac{1+\nu }{8}  \frac{\exp(-x^2)-1 }{x}$ \\[5pt]
		$\widetilde\varepsilon_{rr}(x)$ &   $ \frac{1}{4} (1-2x^2)\exp(-x^2)+ \frac{1+\nu }{8} \frac{1-(1+2x^2)\exp(-x^2)}{x^2} + \frac{1}{2}x^2 \exp(-x^2)$  \\[5pt]
	        $\widetilde\varepsilon_{\theta\theta}(x)$ & $ \frac{1}{4}  \exp(-x^2)+ \frac{1+\nu }{8}  \frac{\exp(-x^2)-1 }{x^2} $\\[5pt] 
	        $ \widetilde p(x)$ & $ \frac{ 1-(1+x^2)\exp(-x^2)}{4 x^2}$   \\[5pt] 
	        $\widetilde{\rho}(x) $ & $ -\frac{2-(2+2x^2+x^4)\exp(-x^2)}{2 x^3}$  \\[5pt] 
	       \hline\hline
  \end{tabular}
\label{tab:Gaussian}
\end{table}
\begin{table}[t]
\caption{Same as Table \ref{tab:Gaussian} but for a parabolic-like bubble profile, as first introduced in Ref.~\cite{Khestanova_NC_2016}. Strain continuity at  $x=1$ implies $c_1=-25/16$ and $c_2=-19(1+\nu)/64$~.}
  \begin{tabular}{ c c }
		\hline\hline
	        Quantity   &   Functional expression \\
	        \hline\hline
	           \\[-8pt]
		$\widetilde h(x)$  &  $\Theta(1-x) (1- \frac{3}{4} x^2 - \frac{1}{4} x^4 )$ \\[5pt]
		$\widetilde u(x)$   & $\left\{
	\begin{array}{ll}
	\frac{1}{192} x \left [398-98 \nu +2 (\nu -7) x^6+12 (\nu -5) x^4+27 (\nu -3) x^2 \right ] +c_1 x  & \mbox{if } x \le 1 \\[5pt]
		c_2 x^{-1} & \mbox{if } x >1
	\end{array}
\right.~.$ \\[10pt]
		$\widetilde\varepsilon_{rr}(x)$ &   $  \left\{
	\begin{array}{ll}
		\frac{1}{192} \left[398-98 \nu +14 (\nu -7) x^6+60 (\nu -5) x^4+81 (\nu -3) x^2\right ]+c_1 + \frac{1}{8} \left(2 x^3+3 x\right)^2& \mbox{if } x \le 1 \\[5pt]
		 -c_2 x^{-2} & \mbox{if } x >1
	\end{array}
\right.$  \\[10pt]
	        $\widetilde\varepsilon_{\theta\theta}(x)$ & $ \left\{
	\begin{array}{ll}
		\frac{1}{192} \left [ 398-98 \nu +2 (\nu -7) x^6+12 (\nu -5) x^4+27 (\nu -3) x^2 \right ] +c_1   & \mbox{if } x \le 1 \\[5pt]
		 c_2 x^{-2} & \mbox{if } x >1
	\end{array}
\right.$\\[10pt] 
	        $ \widetilde p(x)$ & $   \left\{
	\begin{array}{ll}
		-\frac{1}{32} x^2 \left(2 x^4+8 x^2+9\right) & \mbox{if } x \le 1 \\[5pt]
		-\frac{ 19}{32} x^{-2}& \mbox{if } x >1
	\end{array}
\right.$   \\[10pt] 
	        $\widetilde{\rho}(x) $ & $  - \left\{
	\begin{array}{ll}
		-\frac{1}{4} x^3 \left(x^2+2\right)& \mbox{if } x \le 1 \\[5pt]
		\frac{ 19}{8} x^{-3}& \mbox{if } x >1
	\end{array}
\right.$  \\[10pt] 
	       \hline\hline
  \end{tabular}
\label{tab:parabolic}
\end{table}
\begin{table}[t]
\caption{Same as Table \ref{tab:Gaussian} but for a pure-bending bubble profile.  Strain continuity at  $x=1$ implies $c_1=0$ and $c_2=-(1+\nu)/6$. }
  \begin{tabular}{ c c }
		\hline\hline
	        Quantity   &   Functional expression \\
	        \hline\hline
	           \\[-8pt]
		$\widetilde h(x)$  &  $\Theta(1-x) \left ( 1-x^2 \right )^2 $ \\[5pt] 
		$\widetilde u(x)$   & $  \left\{
	\begin{array}{ll}
		\frac{1}{6} x \left [ 4(1- \nu) +(\nu -7) x^6-4 (\nu -5) x^4+6 (\nu -3) x^2\right ]+c_1 x & \mbox{if } x \le 1 \\[5pt]
		 c_2 x^{-1} & \mbox{if } x >1
	\end{array}
\right.$ \\[10pt]
		$\widetilde\varepsilon_{rr}(x)$ &   $\left\{
	\begin{array}{ll}
		\frac{1}{6} \left [ 4(1- \nu) +7 (\nu -7) x^6-20 (\nu -5) x^4+18 (\nu -3) x^2\right ] +c_1 + 8 x^2 \left(1-x^2\right)^2 & \mbox{if } x \le 1 \\[5pt]
		 -c_2 x^{-2} & \mbox{if } x >1
	\end{array}
\right.$  \\[10pt]
	        $\widetilde\varepsilon_{\theta\theta}(x)$ & $\left\{
	\begin{array}{ll}
		\frac{1}{6}\left [ 4(1- \nu) +(\nu -7) x^6-4 (\nu -5) x^4+6 (\nu -3) x^2\right ] +c_1  & \mbox{if } x \le 1 \\[5pt]
		 c_2 x^{-2} & \mbox{if } x >1
	\end{array}
\right. $\\[10pt] 
	        $ \widetilde p(x)$ & $ \left\{
	\begin{array}{ll}
		 -\frac{ 1}{3}   x^2 \left(3 x^4-8 x^2+6\right) & \mbox{if } x \le 1 \\[5pt]
		 -\frac{ 1}{3} x^{-2}& \mbox{if } x >1
	\end{array}
\right.$   \\[10pt] 
	        $\widetilde{\rho}(x) $ & $ 
   \left\{
	\begin{array}{ll}
		-\frac{ 4}{3}  x^3 \left(4-3 x^2\right)& \mbox{if } x \le 1 \\[5pt]
		-\frac{ 4}{3}  x^{-3}& \mbox{if } x >1
	\end{array}
\right.$  \\[10pt] 
	       \hline\hline
  \end{tabular}
\label{tab:bending}
\end{table}
 \begin{figure}[t]
 \centering 
 \begin{overpic}[width=0.48\linewidth]{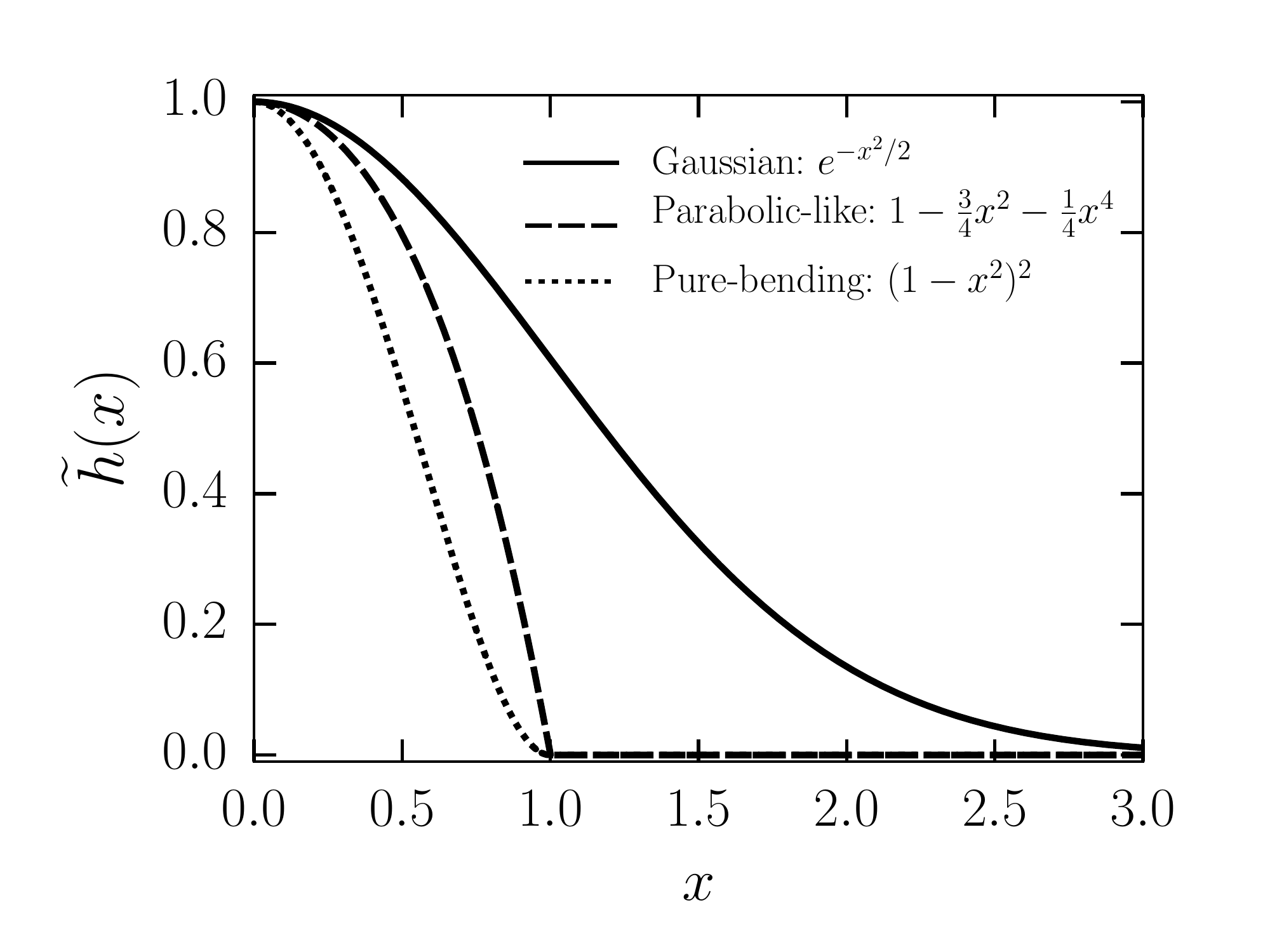}\put(5,66){\large(a)}\end{overpic}
\begin{overpic} [width=0.48\linewidth]{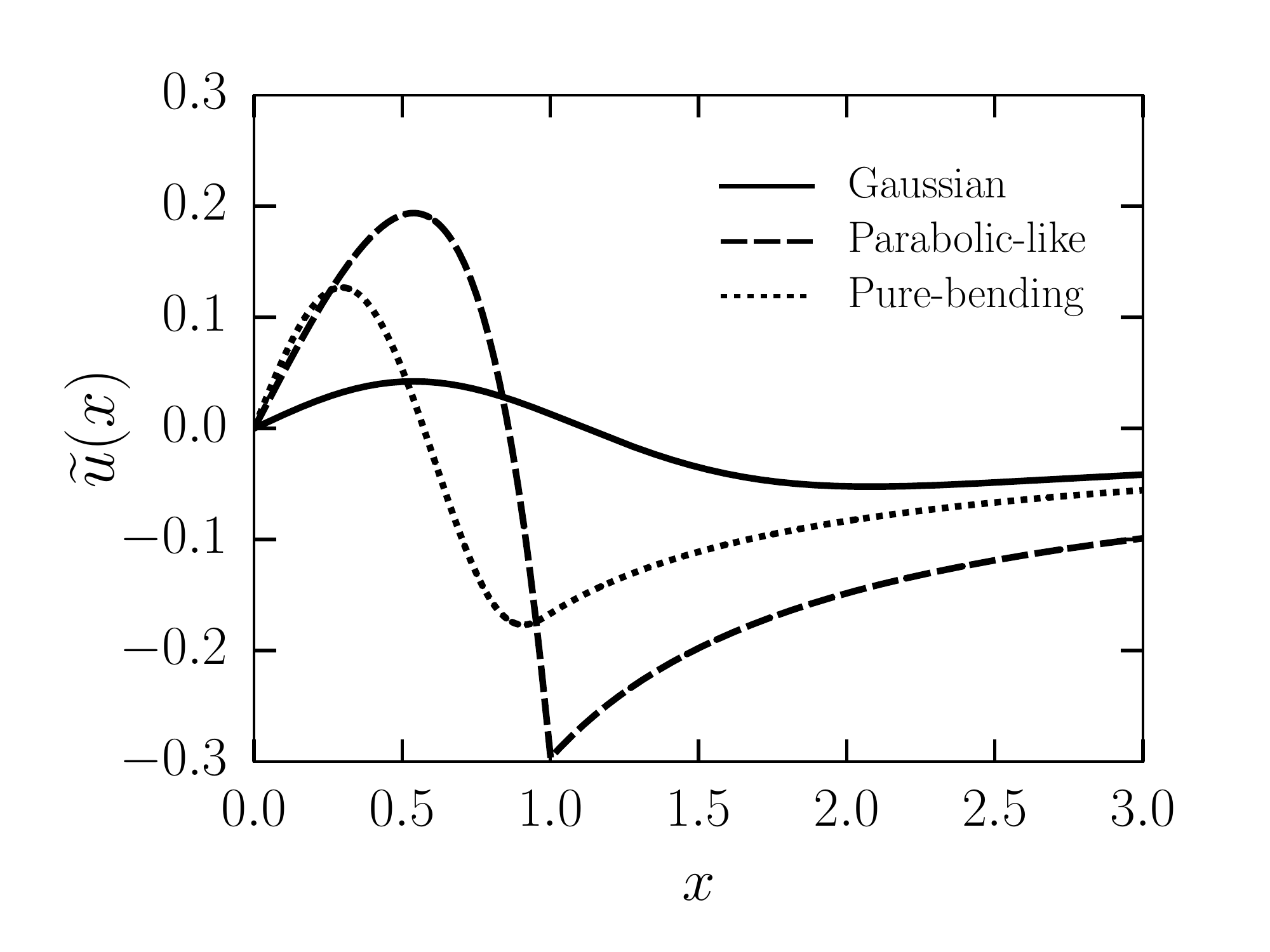}\put(5,66){\large(b)}\end{overpic}
 \begin{overpic}[width=0.48\linewidth]{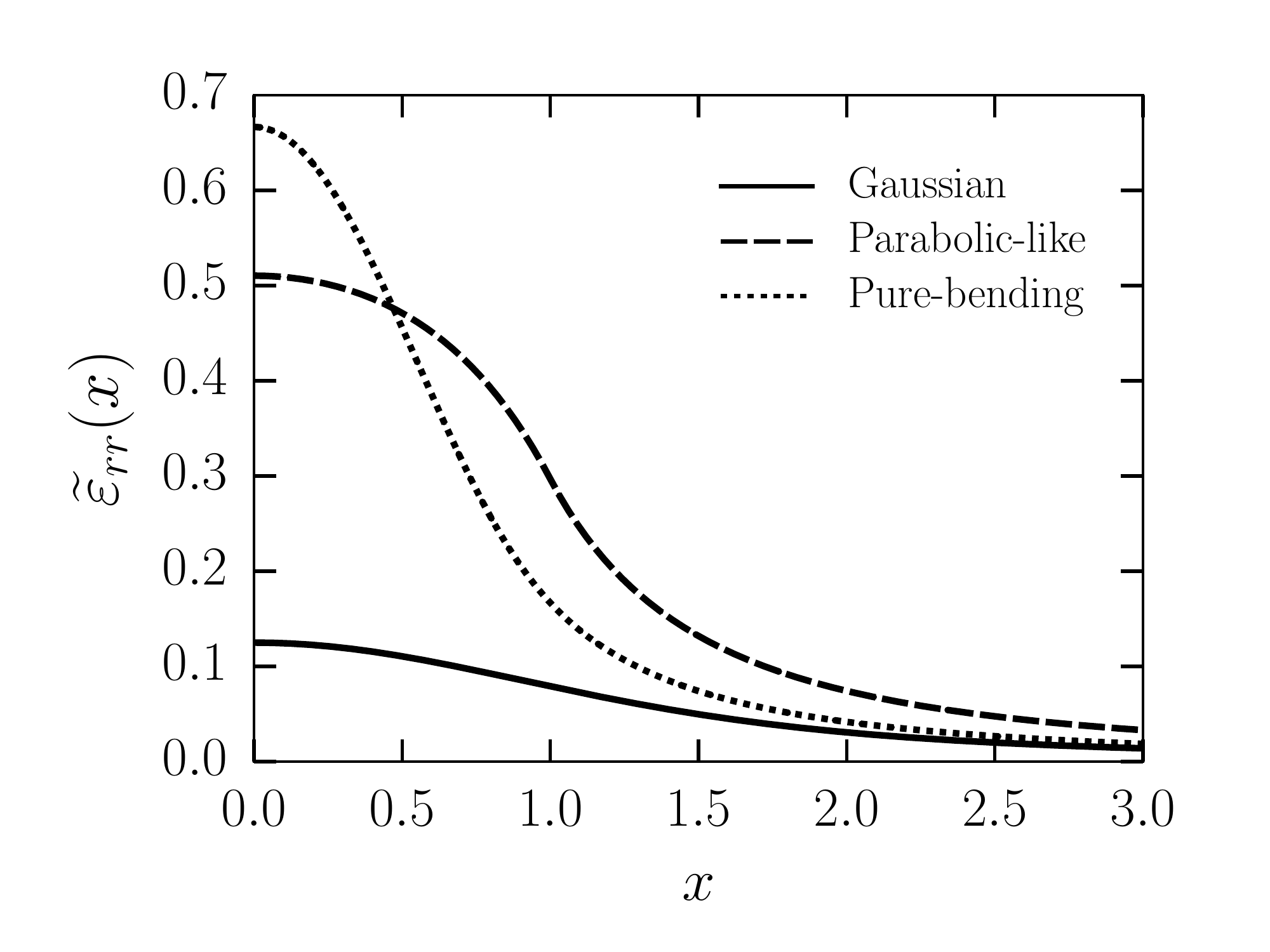}\put(5,66){\large(c)}\end{overpic}
\begin{overpic} [width=0.48\linewidth]{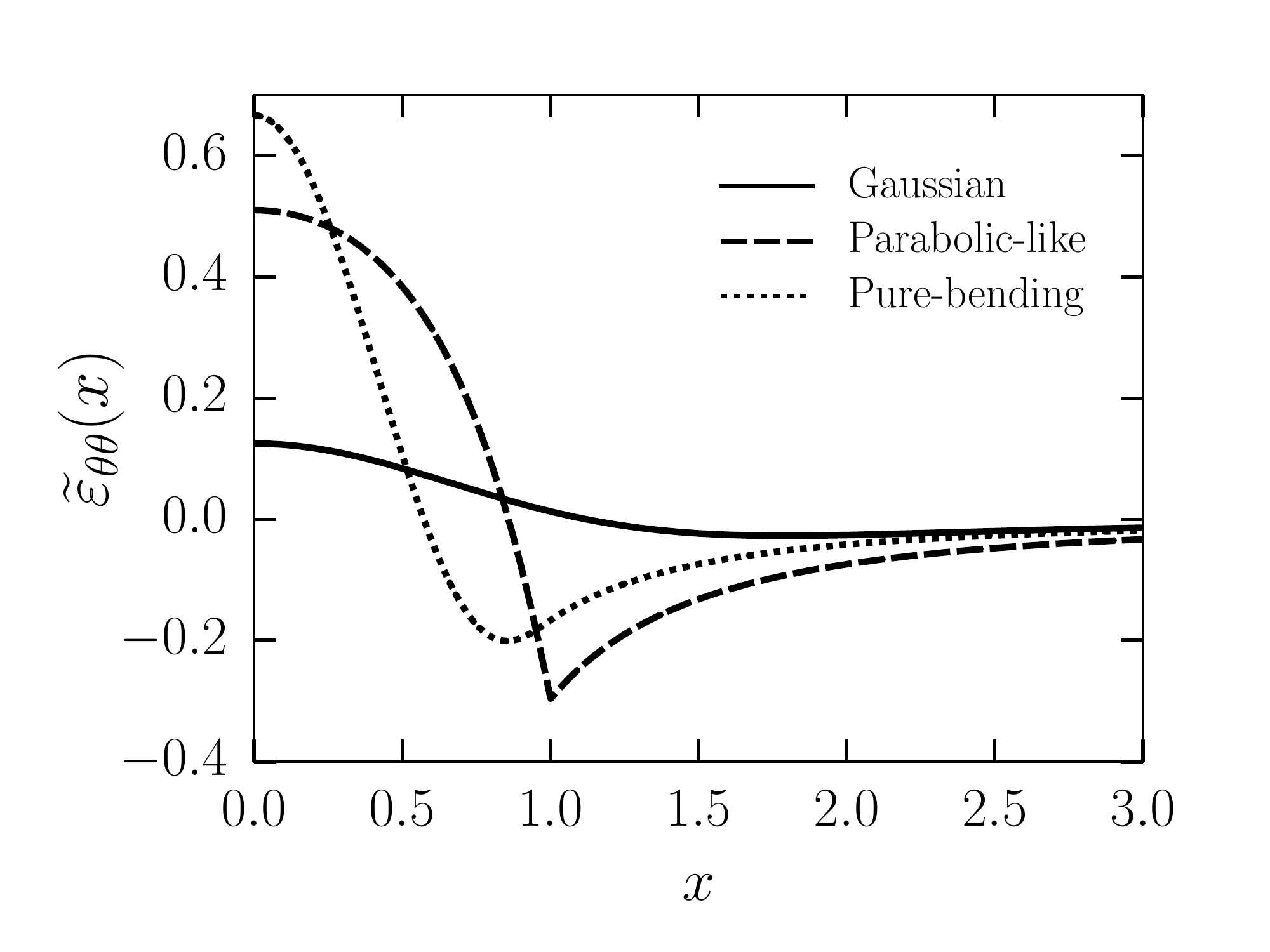}\put(5,66){\large(d)}\end{overpic}
 \begin{overpic}[width=0.48\linewidth]{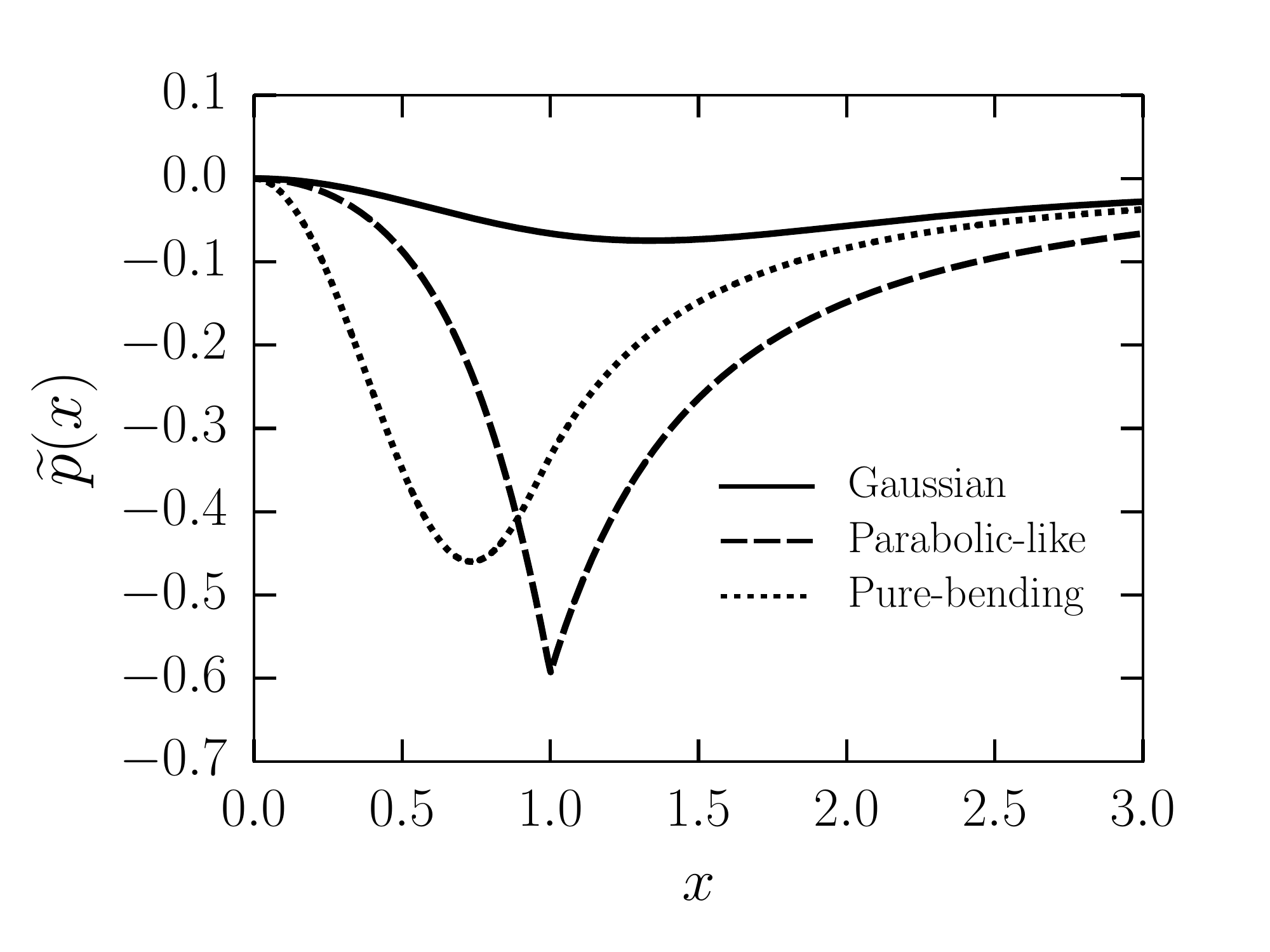}\put(5,66){\large (e)}\end{overpic}
\begin{overpic} [width=0.48\linewidth]{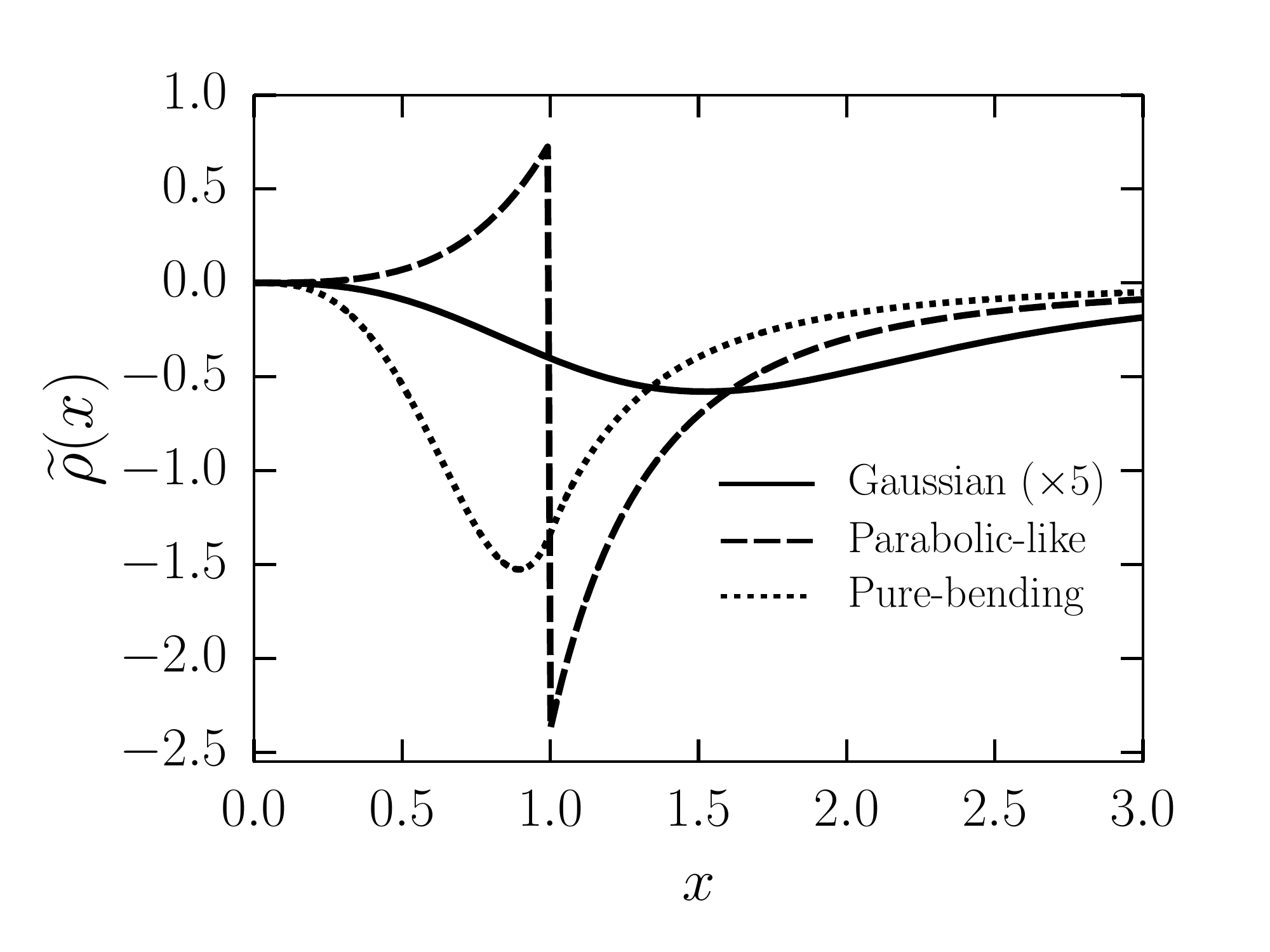}\put(5,66){\large(f)}\end{overpic}
\caption{ 
(a) Out-of-plane displacement, $\widetilde h(x)$. 
(b) Radial displacement, $\widetilde u(x)$.  
(c) Radial strain, $\widetilde \varepsilon_{rr}(x)$. 
(d) Azimuthal strain, $\widetilde \varepsilon_{\theta\theta}(x)$.
(e) The radial profiles of polarization, $\widetilde p(x)$. 
(f) The radial profiles of induced charge density, $\widetilde \rho(x)$.
We set $\nu=0$ in these plots.}
\label{fig}
\end{figure}
 \end{widetext}
\end{document}